\documentclass[useAMS,usenatbib,letter]{mn2e}
\usepackage{aas_macros}
\usepackage[dvips]{graphicx}
\usepackage{amssymb}
\title[Orientations of Disk Galaxies]{The Large-Scale Orientations of Disk Galaxies}
\author[O. Hahn et al.]{Oliver Hahn$^{1,2}$\thanks{E-mail: hahn@phys.ethz.ch},
Romain Teyssier$^{3,4}$ and C. Marcella Carollo$^{1}$ \\
$^{1}$Department of Physics, ETH Zurich, CH-8093 Z\"urich, Switzerland\\
$^{2}$KIPAC, Stanford University, 2575 Sand Hill Road, Menlo Park, CA 94025, USA\\
$^{3}$Institute for Theoretical Physics, University of Zurich, CH-8057 Z\"urich, Switzerland\\
$^{4}$CEA Saclay, DSM/IRFU/SAP, B\^atiment 709, F-91191 Gif-sur-Yvette, Cedex, France}
\begin{document}

\date{draft v.2}
\pagerange{\pageref{firstpage}--\pageref{lastpage}} \pubyear{2010}
\maketitle

\label{firstpage}

\begin{abstract}
We use a $380\,h^{-1}{\rm pc}$ resolution hydrodynamic AMR simulation of a cosmic filament to investigate the orientations
of a sample of $\sim100$ well-resolved galactic disks spanning two orders of magnitude in both stellar
and halo mass.  We find: (i) At $z=0$, there is an almost perfect alignment  at a median  angle of 
$18^\circ$, in the inner dark matter halo regions where the disks reside, 
between the spin vector of the gaseous and stellar galactic disks and that of their inner host haloes. 
The alignment between  galaxy spin and  spin of the entire host halo is however 
significantly weaker, ranging from a median of $\sim46^\circ$ at $z=1$ to $\sim50^\circ$ at $z=0$. (ii)
The most massive galaxy disks have spins preferentially aligned so as to point along their host  filaments. 
(iii) The spin of  disks in lower-mass haloes shows, at redshifts above $z\sim0.5$ and in regions
of low environmental density,  a clear signature of alignment with the intermediate principal axis of the 
large-scale tidal field. This behavior is consistent with  predictions of  linear tidal torque theory. This 
alignment decreases with increasing environmental density, and vanishes in the highest density regions. 
Non-linear effects in the high density environments  are plausibly responsible for establishing 
this density-alignment correlation. We expect that our numerical results provide important insights
for both understanding intrinsic alignment in weak lensing from the astrophysical perspective
and formation and evolution processes of galactic disks in a cosmological context.
\end{abstract}



\section{Introduction}
Numerical simulations and analytic calculations have shown that the gravitational
growth of tiny density perturbations in the $\Lambda$CDM cosmological model
leads to a wealth of structures over cosmic time.
The spatial distribution of gravitationally bound structures follows a web-like pattern
composed of dense clusters, filaments, sheets and otherwise near-empty void regions
\citep[see e.g.][]{Shandarin1989,Bond1996}. This ``cosmic web'' is the weakly nonlinear manifestation of the 
large-scale tidal field and its formation can be readily understood in the first order
Lagrangian perturbation theory of the Zel'dovich approximation \citep{Zeldovich70}. 
Recent large scale galaxy surveys have observationally confirmed the presence of the cosmic web in the Universe
\citep[e.g.][]{Colless2001, Tegmark2004}. 

Since galaxies form and evolve together with the cosmic web in which they are embedded,
it is a question of high relevance whether the properties of galaxies depend
on scales beyond the most immediate vicinity and thus reflect the effect of large-scale
tidal fields on the assembly of the galaxy. General ellipsoidal density perturbations collapse
sequentially along their three axes leading to matter accreting first onto a sheet, then
collapsing to a filament before matter finally streams along the filaments into the densest
regions. As a consequence of these processes, the large-scale morphology of the cosmic web might leave
its imprint on the orientations of galaxies through the anisotropy of accretion and mergers.

Recently, evidence has accumulated that galaxies are indeed not oriented randomly
with respect to one another. On small scales, several studies report a radial alignment of satellite galaxies 
towards the centre of mass of the group or cluster in which they reside 
\citep[e.g.][]{Yang2006,Faltenbacher2007,Faltenbacher2008}. 
Such an alignment is predicted by N-body 
simulations of dark matter halos \citep[e.g.][]{Bailin2005a,Pereira2008,Knebe2008}. It is commonly 
believed that tidal torques are responsible for this radial alignment \citep[e.g.][]{Pereira2008}, 
as tidal forces on these relatively small scales ($\lesssim1$ Mpc) are sufficiently strong to affect halo 
shapes on short time-scales.

More surprisingly, however, observational results indicate that the orientations of galaxies  
are correlated up to scales of order $\sim100$ times 
larger than the virial radii of their host haloes. \cite{Mandelbaum2006}, \cite{Okumura2009}
 and \cite{Faltenbacher2009} detect an alignment of 
galaxy shapes for the luminous red galaxies (LRGs) in the SDSS survey out to the largest probed 
scales ($\sim80$ Mpc). \cite{Hirata2007} report a similar result for redshifts $0.4<z<0.8$. 
For lower luminosity galaxies, in particular for $L<L_\ast$, no alignment has been detected 
yet \cite[see also][]{Slosar2009} although some results using N-body simulations indicate
spin alignment around voids \citep{Brunino2007,Cuesta2008}. Correlations of the spin of galaxies have been the subject
of several analytical models \citep[see e.g.][for a review, and references therein]{Schaefer2009}.

Compared with more 
local effects, the relative weakness on galaxy scales of tidal forces arising from the $\gg1 {\rm Mpc}$ scales requires 
that such effects maintain spatial coherence over rather long time scales for them to be able to affect 
galaxy properties. Also it is not clear how they are counteracted by local non-linear
effects in the most immediate vicinity of the galaxy: non-linear gravitational effects as well as ram pressure
exerted on the disk gas could potentially lead to a reorientation of galactic angular momentum.
 Furthermore, the impact of hot in contrast to cold mode accretion separated by
a halo mass of $\sim4\times10^{11}\,h^{-1}{\rm M}_\odot$ of gas onto galaxies
\citep[cf.][]{Birnboim2003,Keres2005,Dekel2006,Ocvirk2008} on the orientation of the resulting
disks is not clear.

While understanding the formation of galaxies in the context of their host large-scale structure is a key
question for galaxy formation theory, a precise knowledge of correlated galaxy orientations is of high
importance also for future high-precision weak lensing studies of dark energy. Weak lensing signals 
can be severely affected by systematic contaminations due to large-scale alignments in 
the spatial distribution of cosmic matter on scales from a few to more than a hundred Mpc (due to galaxy 
ellipticity-ellipticity alignment: \cite{Catelan2001}; and due to the much more problematic shear-ellipticity 
alignment: \cite{Hirata2004}). Approaches that have been suggested so far to remove the contamination due 
to shear-ellipticity alignment suffer from several limitations and uncertainties: the statistical down-weighting of 
the effect proposed by \cite{Joachimi2008} relies on precision determinations of photometric 
galaxy redshifts; other techniques, such as proposed by e.g. \cite{King2005}, depend heavily on toy models 
\citep[e.g.][]{Catelan2001} which do not include any (as yet unknown) dependence of alignments on 
galaxy luminosity, mass, or redshift. Several studies using $N$-body simulations however find evidence for a dependence of 
the strength of halo-LSS alignment on mass and redshift \citep[see e.g.][]{Hahn2007b,Aragon06,Paz2008} . 

Unfortunately, it is currently still computationally unfeasible to use hydrodynamical simulations of galaxy
formation of the necessary resolution in a sufficiently large volume to investigate the correlation function
of galaxy shapes and spins. To circumvent this barrier, some authors have used large-volume $N$-body
simulations together with a prescription to assign galactic disk orientations to the dark matter haloes
which typically either assumes perfect alignment with the DM halo spin or some random misalignment 
angle between the two \citep[e.g.][]{Heymans2006}. Also analytic models based on the halo model
have been developed that include prescriptions for small scale satellite-central
alignments \citep{Schneider2009}. Probability distributions for the angles between halo and galaxy
spins have been determined by various authors \citep[e.g.][]{vdBosch2002,Bailin2005b,Sharma2005,Croft2008,Bett2009}.

Accepting the current unfeasibility of large volume hydrodynamic simulations, we propose a different approach.
Combining local measurements of the alignment of galaxies with the large-scale structure -- as has been
investigated by \cite{Hahn2007a,Hahn2007b} and \cite{Aragon06} in pure $N$-body simulations --
circumvents the need for large volumes but can still provide constraints on the degree of galaxy
alignment to be expected. The results of \cite{Navarro04} on the basis of four SPH simulations of the formation
of single isolated galaxies indicate an alignment of the spin of these galaxies with the intermediate principal
axis of the tidal field at turnaround that persists to a slightly weaker degree also to $z=0$.

In this paper, we study the orientation of a sample of $\sim100$ galactic discs with respect to their larger scale
environment in a hydrodynamical cosmological simulation of the formation of a large filament.
We have chosen this particular large-scale environment as it provides an environment where
the gravitational shear is expected to be relatively large so that weak lensing shear should
have good signal-to-noise. Any intrinsic alignment of galaxies on these scales will thus
impact the measurements through the GI-lensing term. Prior investigations in dark matter
cosmological simulations have shown evidence for alignment of halo spin with 
filaments. However, considering the gas component rather than dark matter, the filament 
is a high pressure environment so that hydrodynamic effects, such as ram-pressure, 
are expected to play a role in the evolution of the simulated galaxies.

This paper is organised as follows: we describe the details of our numerical simulation, the method to identify
galaxies and the selection of galaxies for our analysis in Section \ref{sec:galform_methods}. We first study
the alignments between the gas, stellar, dark matter angular momenta and the total halo spin in Section 
\ref{sec:galform_aligninternal}. Then, we investigate in Section \ref{sec:galform_align_lss} the alignment of the 
various components with the cosmic large-scale structure, quantitatively described by the tidal field.
In Section \ref{sec:numerical_artefacts} we discuss the numerical convergence of our results, before
a comparison of our results to observations is given in Section \ref{sec:galform_observations}.
Finally, we summarise and provide our conclusions in Section \ref{sec:galform_summary}.


\section{Numerical Methods}
\label{sec:galform_methods}

\begin{figure*}
  \begin{center}
    \includegraphics[width=0.4\textwidth]{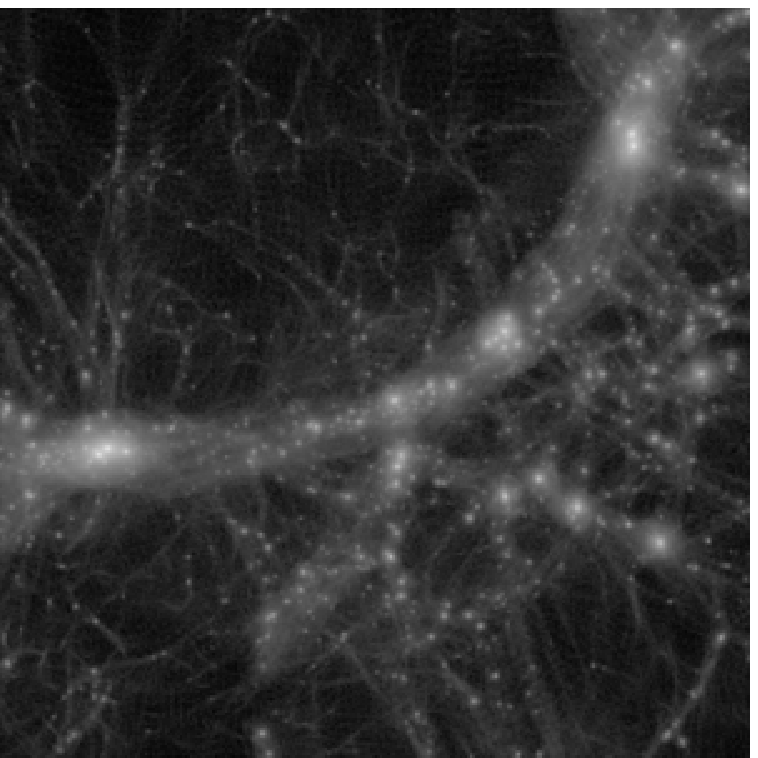}
    \hspace{7mm}
    \includegraphics[width=0.4\textwidth]{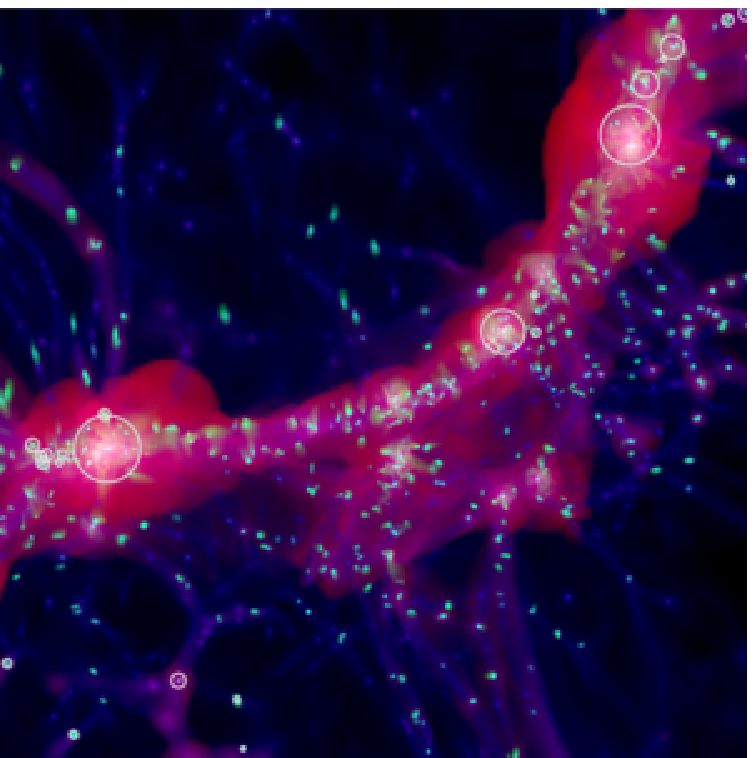}
  \end{center}
  
  \caption{ \label{fig:TheFilament}The simulated filament at redshift $z=0$. The images extend over
  $20\,h^{-1}{\rm Mpc}\times20\,h^{-1}{\rm Mpc}$ and have a projection depth of $10\,h^{-1}{\rm Mpc}$.
  (Left panel:) the dark matter
  particle distribution, (Right panel:) a composite RGB colour image of the gas distribution, the red image
  channel was assigned to the pressure, the green channel to the metallicity of the gas and the
  blue channel to the density. Since metal enriched gas is produced only in galaxies, each of the 
  green blobs corresponds to a galaxy. The white circles in the right panel indicate the virial radii of those host haloes
  within the filament that have been excluded from refinement for most part of the simulation or are too close
  to a low-resolution region in the high resolution run (cf. Section \ref{sec:filament_sim}) and are thus not 
  considered in our analysis.  All other galaxies are treated with full resolution and are not affected by low
  resolution outside of the zoom region.}
\end{figure*}

\begin{figure}
  \begin{center}
  \includegraphics[width=0.35\textwidth]{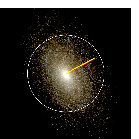}\\
  \includegraphics[width=0.35\textwidth]{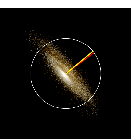}  
  \end{center}
  
  \caption{ \label{fig:TheGalaxies} Simulated stellar light composite images of two well-resolved disk-galaxies 
  using the $i'$, $r'$ and $g'$ band light and taking into account dust absorption. The white circles 
  correspond to the radius containing 95 per cent of the stellar mass. The yellow line corresponds
  to the stellar angular momentum vector direction, the red line to the gas angular momentum.}
\end{figure}

\subsection{Specifics of the Cosmological Simulation}
\label{sec:filament_sim}
We simulate the formation of a large cosmic filament and the galaxies embedded in it
using the AMR code {\sc Ramses} \citep{Teyssier2002} to evolve dark matter, gas
and stars. Gas dynamics is computed using a second-order unsplit Godunov scheme,
particles (dark matter and stars) are evolved using the particle-mesh method.
The simulation includes standard recipes for star formation \citep{Rasera2006},
supernova feedback and chemical enrichment \citep{Dubois2008}. We do not 
repeat details of these sub-grid models but kindly refer the reader to the mentioned
papers. The filament was chosen by visual inspection of the environment of the
most massive haloes in a dark-matter-only preflight simulation. We then chose
a filament that is very prominent (and thus relatively massive) and not criss-crossed by 
too many smaller scale filaments. 

Initial conditions were generated using the {\sc Grafic-2} tool
from a power spectrum computed with {\sc Linger}. We adopt  
cosmological parameters as obtained from the WMAP 5th year data release
\citep{Komatsu2009}, i.e. we use the density parameters $\Omega_m=0.276$, $\Omega_b=0.046$,
$\Omega_\Lambda=0.724$, a power spectrum normalisation of $\sigma_8=0.811$ and
a spectral index $n_s=0.961$ for the primordial spectrum, 
as well as a Hubble constant of $H_0 = 100\, h\, {\rm km}\,{\rm s}^{-1}\,{\rm Mpc}^{-1}$
with $h=0.703$.

The filament is simulated using the ``zoom-in'' technique as a high-resolution region of $~25\,h^{-1}{\rm Mpc}$
diameter at redshift $z=0$ inside of a comoving $100\,h^{-1}\textrm{Mpc}$ cosmological box.
 The effective initial resolution of our simulation
is $1024^3$ particles and grid points inside this high-resolution region, dropping in
steps of 8 to $128^3$ in the remainder of the box. Additional refinements are confined
to the high-resolution region only. We adopt a refinement strategy based on density 
threshold so that the number of particles per cell (and the baryon mass per cell) remains roughly constant
in time. Since the filament moves slightly in space, we introduce a ``colour'' field
to mark the 
 Lagrangian volume of the filament. This field is passively advected
during the formation of the structures and thus follows the evolution of the filament perfectly.
In order to save memory, we used as colouring variable the metallicity of the gas, with a very low 
(but non-zero)
value of $10^{-3}$ solar in the high resolution region and exactly zero outside. 
Additional refinements are only allowed where the metallicity
exceeds one tenth of this initial value. In addition, cooling and star formation are also 
switched off in non-coloured regions. The maximum resolution of our simulation is 
{\it physical} $0.38\,h^{-1}\textrm{kpc}$ 
at all times. In addition, our Lagrangian volume excludes the most massive galaxy groups
(halo mass above $5\times 10^{13}\,h^{-1}{\rm M}_\odot$ at $z=0$)
in order to speed up computation as we are only interested in smaller
objects in this study. 

The two panels of Figure \ref{fig:TheFilament} give a visual impression of the filament at $z=0$.
We show two panels, the first illustrating the distribution of dark matter, the second is a 
composite image representing the density, pressure and metallicity distribution
in the gas. Note that the main filament is entirely wrapped by an
accretion shock. It is thus clearly a high pressure environment. The less massive
filaments remain cold except at node points where the more massive haloes reside.
An inspection of the time-evolution of the simulation reveals that this accretion shock emerges
first from the most massive haloes along the filament but then grows along the 
filament boundaries until the entire filament is enclosed by a shock.

In the two panels of Figure \ref{fig:TheGalaxies}, we show two examples of well-resolved
disk-galaxies at $z=0$ in the simulation. The images represent a simulated stellar light
composite image combining the i', r' and g' bands into an RGB image taking into 
account dust reddening along the line of sight. The images also show the orientation
of the stellar and gaseous angular momentum and indicate that these are perpendicular
to the disks and well aligned with one another.

\subsection{\label{sec:galfind}Identification of Galaxies}
In order to identify galaxies in the simulated data, we directly use the AMR mesh
structure and label contiguous volumes exceeding a density threshold.  
First, we pick the highest refinement level which exceeds
$1 h^{-1}$kpc resolution in order to smooth out galactic substructure. Next,
we keep only those cells on this level, which exceed a threshold $\delta_b=50000$ 
in baryonic overdensity. This particular choice of $\delta_b$ has been found to be
in excellent agreement with a visual identification of the galaxies.
In a final step we label all contiguous regions that
are left. This labelling is performed using a parallel implementation of the 
Hoshen-Kopelman cluster finding algorithm \citep{Hoshen1976}. Finally, 
all star particles and gas cells (also on higher refinement levels) are assigned
to the mesh cluster $\mathcal{G}_k$, i.e. the ``galaxy'',  in which they reside.

Galaxy stellar centres are then determined by iteratively computing the stellar centre of mass
within shrinking spherical apertures and moving the aperture to the new centre
in each iteration. The procedure is stopped once less than 50 particles are 
contained within the aperture. The centre of the gaseous disk is defined as
the densest gas cell in the galaxy and the centre of the inner dark matter halo
is defined as the most bound dark matter particle when evaluating the potential
for all dark matter particles that are contained in the galaxy patch.

In order to determine halo masses and virial radii, we grow spheres around each galaxy 
centre until the enclosed matter overdensity falls below the overdensity $\Delta_c$ of a 
virialised spherical perturbation in a $\Lambda$CDM cosmology as given by the fitting
formula of \cite{Bryan1998}. In a next step, we determine equivalence classes
of galaxies that are contained within the virial radii of each other. The galaxy
with the highest stellar mass in each equivalence class we call the ``central
galaxy'', all other class members are ``satellite galaxies''. We do not consider
these satellite galaxies further in this study. 

Finally, unresolved galaxies are identified by tracking the dark matter particles 
that end up in galaxies back to the initial conditions and requiring that none of
them started in a region that was not initially marked for cooling and refinement.
These unresolved galaxies are then removed from the galaxy catalogue.

The central galaxies in our samples have stellar radii $r_{95}$ (which is the radius that contains 95 per cent
of the stellar mass) that are at most 0.11 at $z=1$, 0.095 at $z=0.5$ and $0.077$ at $z=0$ of the virial 
radius of their parent halo. This makes us confident that the identified galaxies are not contaminated 
by other satellites in the halo and thus constitute clean samples of central galaxies.

\subsection{\label{sec:galsamples}The Fiducial Galaxy Samples}
For our analysis, we keep only the best resolved galaxies with at least 25000 star particles,
corresponding to a minimum stellar mass of $\sim 7.5\times 10^{9}h^{-1}{\rm M}_\odot$.
For these galaxies, $r_{80}$, the radius containing 80\% of the stellar mass is
resolved by at least 8 resolution elements. Furthermore, this requirement implies that we can
compare our results to a lower resolution run (see Appendix \ref{sec:numerical_artefacts})
for which the minimum star particle
number for galaxies of equal mass is then a factor of 8 lower, but still above 3000.

\begin{figure*}
  \begin{center}
    \includegraphics[width=0.67\textwidth]{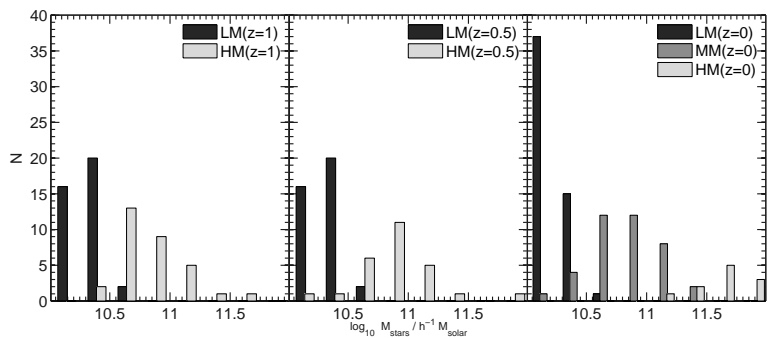}
  \end{center}
  \caption{ \label{fig:MassDistributions} Distributions of galaxy stellar masses in the two/three galaxy samples
  used in our analysis (see Table \ref{tab:halo_samples} for details) at redshifts $z=1$ (left), 
  $z=0.5$ (middle) and $z=0$ (right).}
\end{figure*}

We split the sample of  well-resolved galaxies according to their host halo masses into 3
different halo groups. The high-mass sample is defined as all halos above the 
non-linear mass $M_\ast(z=0)\approx 2\times10^{12}h^{-1}{\rm M}_\odot$.
Prior results using pure $N$-body simulations have
indicated that $M_\ast$ might be a characteristic scale for the spin-alignment behaviour
of dark matter haloes \citep[see e.g.][]{Hahn2007b, Aragon06, Paz2008}. The intermediate sample is
for halo masses smaller than this non--linear mass, but larger than the ``cold accretion limit'' 
of $\sim 4\times 10^{11}h^{-1}{\rm M}_\odot$ \citep[see e.g.][]{Birnboim2003,Keres2005,Dekel2006}.
Haloes above this critical mass are believed to host a hot pressurized atmosphere that can exert
ram pressure effects on the central galaxy. The third mass sample is for halo masses smaller than 
the critical mass. Note that for the two higher redshifts $z=1$ and $z=0.5$, we merge together
the high and intermediate mass samples to improve statistical significance.
Detailed information about the galaxy samples is given in Table \ref{tab:halo_samples}. 
The distribution of stellar masses in the samples is shown in Figure \ref{fig:MassDistributions}.

\begin{table*}
\begin{center}

\begin{tabular}{rccc}
\hline
 & LM & MM & HM \\
\hline
\multicolumn{1}{l}{$z=0$:} \\
$M_{\rm halo}$ / $h^{-1}{\rm M}_\odot$ & $1-4\times10^{11}$ & $5\times10^{11}-2\times10^{12}$ & $ >2\times10^{12}$\\
median $M_{\rm stellar}$  / $h^{-1}{\rm M}_\odot$ & $1.3\times10^{10}$ & $6.8\times10^{10}$ & $4.2\times10^{11}$ \\
number of galaxies & 70 & 39 & 12 \\
\hline
\multicolumn{1}{l}{$z=0.5$:} \\
$M_{\rm halo}$ / $h^{-1}{\rm M}_\odot$ & $1-4\times10^{11}$ & - & $ >4\times10^{11}$\\
median $M_{\rm stellar}$ / $h^{-1}{\rm M}_\odot$ & $1.3\times10^{10}$ & - & $6.7\times10^{10}$ \\
number of galaxies  & 48 & - & 28 \\
\hline
\multicolumn{1}{l}{$z=1$:} \\
$M_{\rm halo}$ / $h^{-1}{\rm M}_\odot$ & $1-4\times10^{11}$ & - & $ >4\times10^{11}$\\
median $M_{\rm stellar}$ / $h^{-1}{\rm M}_\odot$ & $1.3\times10^{10}$ & - & $6.3\times10^{10}$\\
number of galaxies & 58 & - & 31 \\
\hline
\end{tabular}
\vspace{0.5cm}
\end{center}
\caption{\label{tab:halo_samples}The galaxy samples used in our analysis. We split central galaxies according
to their host halo mass in a low mass (LM) and high mass (HM) bin at redshifts $z=1$ and $z=0.5$ and an additional
medium mass (MM) bin at $z=0$. For each mass bin and each redshift, the median stellar mass as well as the
number of galaxies in the bin is given. See Figure \ref{fig:MassDistributions} for histograms of stellar masses for each of
the galaxy samples.}
\end{table*}

\subsection{Measured Galactic Properties}
In this study, we are interested in the orientation of disk galaxies. Since we exclude the most massive
(large group-scale and above) haloes from refinement, indeed all galaxies considered are disky. 
We can thus simply quantify the disk orientation by the angular momenta of the various disk 
constituents: gas, stars and dark matter.

We compute the angular momentum for each galaxy $\mathcal{G}_k$ by evaluating the sum
\begin{equation}
{\bf J}_k = \sum_{i\in \mathcal{G}_k} m_i({\bf r}_i - \bar{{\bf r}}_k)\times({\bf v}_i - \bar{{\bf v}}_k),
\label{eq:angmom}
\end{equation}
where $\bar{\bf r}_k$ is the stellar centre of the $k$th galaxy inner region (see previous section), 
and $\bar{\bf v}_k$ is the centre of mass velocity. The sum is evaluated separately for all star particles that are part of an
identified galaxy $\mathcal{G}_k$ to yield the stellar angular momentum ${\bf J}_k^{\rm S}$, 
as well as for all DM particles that are contained within a sphere of the radius of the most 
distant star particle in the galaxy ${\bf J}_k^{\rm DM}$. Finally, we also compute the
gas angular momentum ${\bf J}_k^{\rm G}$ by summing over all leaf cells of the AMR
tree that are contained in the galaxy (and are thus effectively bounded by an iso-density 
surface -- see Section \ref{sec:galfind} for details).  For gas, $\bar{\bf r}_k$ refers to the densest
cell in the galaxy patch and $\bar{{\bf v}}_k$ to the centre of mass velocity computed over all
leaf cells in $\mathcal{G}_k$, while for dark matter $\bar{\bf r}_k$ refers to the most bound dark matter particle
and $\bar{\bf v}_k$ to the centre of mass velocity of the dark matter particles in the sphere
containing $\mathcal{G}_k$ (see also Section \ref{sec:galfind}). In order to determine the total halo angular momentum, 
we evaluate eq. (\ref{eq:angmom}) summing over all components and up to the virial 
radius to obtain ${\bf J}_{\rm tot}$.


\subsection{Quantifying Large-Scale Structure with the Tidal Field}

\begin{figure*}
  \begin{center}
    \includegraphics[width=0.8\textwidth]{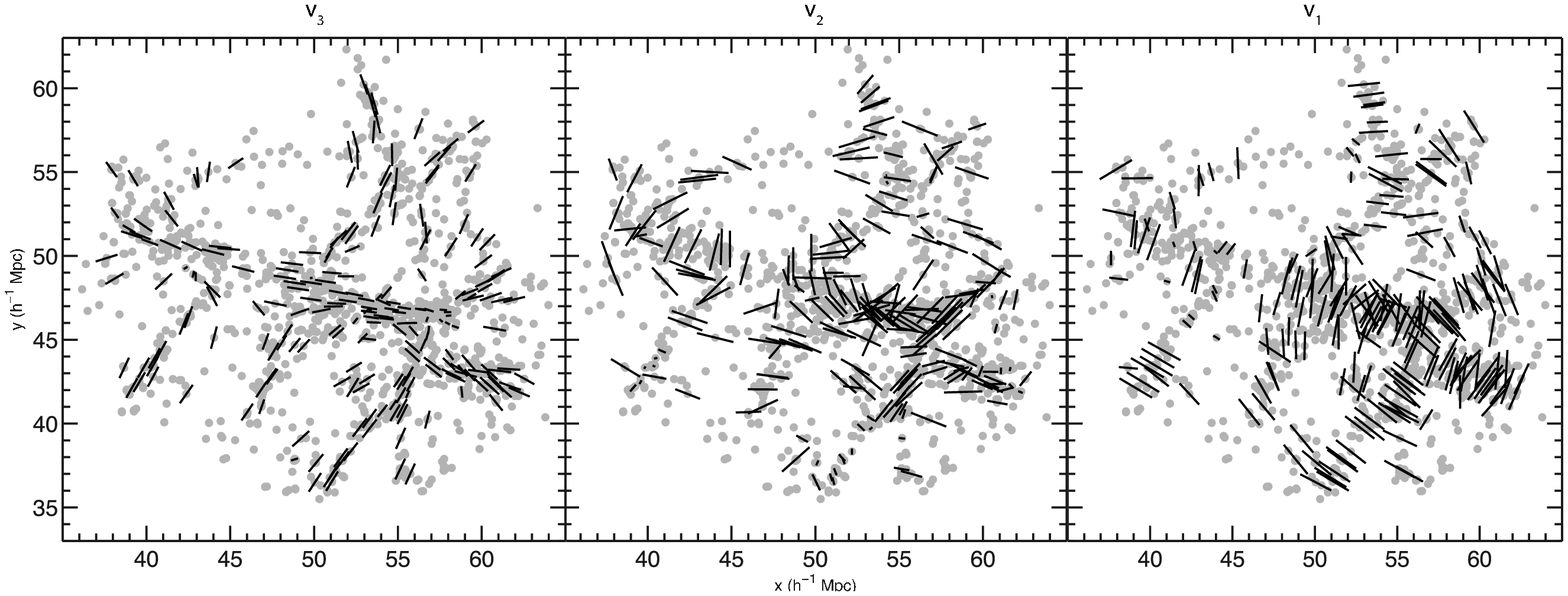}
  \end{center}
  \caption{ \label{fig:quiver_web} Directions of the tidal field eigenvectors at $z=0.5$ for a smoothing
  scale of $R_s=2\,h^{-1}{\rm Mpc}$ physical ($3\,h^{-1}{\rm Mpc}$ comoving). The ${\bf v}_3$
  eigenvector indicates the overall direction of the filaments, while ${\bf v}_2$ and ${\bf v}_1$
  span the perpendicular plane. Linear tidal torque theory favours alignment of galactic spin
  with ${\bf v}_2$.}
\end{figure*}

In analogy to the approach followed by \cite{Hahn2007a,Hahn2007b}, we
use the eigenvectors of the tidal field tensor smoothed on some scale $R_s$ to probe 
alignments with the large-scale structure. This approach has two main advantages:
(1) as demonstrated in \cite{Hahn2007a}, the tidal field provides a natural dynamical
quantification of the geometry of the large-scale structure; and (2) any local alignments with the tidal
field directly translate into shear-ellipticity alignments in weak lensing \citep{Hirata2004}. In
addition, when supplemented with auto-correlation functions of the non-linear tidal
field \cite[see][]{Lee2009a}, local alignments with the tidal field can be translated into
correlation functions. We will investigate this approach in future work. 

We determine the tidal field by first constructing the full matter overdensity field $\delta({\bf x})$
of the entire simulation volume on a $512^3$ grid. This grid is subsequently smoothed with a
Gaussian kernel of varying filtering radius $R_s$. Finally, we solve Poisson's equation
using the fast Fourier transform (FFT) to obtain the potential $\phi({\bf x})$, evaluate the
Hessian matrix $T_{ij}\equiv\partial_i\partial_j \phi$ by finite differencing the potential and interpolating
via cloud-in-cell to the galaxy positions. Diagonalising $T_{ij}$ at each galaxy's position,
we find the three eigenvalues $\lambda_3{(R_s)}\leq\lambda_2{(R_s)}\leq\lambda_1{(R_s)}$
and the corresponding unit eigenvectors ${\bf v}_3{(R_s)},{\bf v}_2{(R_s)},{\bf v}_1{(R_s)}$
as a function of the applied filtering scale $R_s$. Note that we define $T_{ij}$ to be simply the
Hessian of the potential, i.e. the signs of the eigenvalues might be reversed with respect to
some other studies.

The eigenstructure of the Hessian of the gravitational potential is intimately connected to
the large-scale dynamics of the cosmic web. Filamentary regions tend to have one negative
and two positive eigenvalues when smoothed on large-enough scales. In particular, we find
e.g. average values $\left<\lambda_3\right>=-0.2$, $\left<\lambda_2\right>=0.2$ and 
$\left<\lambda_1\right>=0.6$ on scales of comoving $3\,h^{-1}{\rm Mpc}$ where the averages are
taken over the sample of our galaxies at $z=0.5$. Units are such that $\sum_i\lambda_i=\delta$,
where $\delta$ is the matter overdensity. The signature of these eigenvalues indicates that the galaxies 
are, on average, indeed embedded in a filamentary environment. The signs of the eigenvalues reflect 
the large-scale expansion along the filament -- corresponding to the direction indicated by the
eigenvector of the negative eigenvalue $\lambda_3$ -- and gravitational contraction in the
perpendicular plane -- corresponding to the plane spanned by the eigenvectors ${\bf v}_2$
and ${\bf v}_1$. The connection between the eigenstructure of $T_{ij}$ and dynamics follows
readily from the Zel'dovich theory of pancake collapse \citep{Zeldovich70}. In Figure \ref{fig:quiver_web},
we show the orientations of the eigenvectors with respect to the cosmic web structure.

To summarize, ${\bf v}_3$ indicates the overall direction of a filament, while ${\bf v}_2$ and ${\bf v}_1$
span the perpendicular plane. A more detailed account is given in \cite{Hahn2007a}.


\section{The Distribution of Angular Momentum in Galaxies}
\label{sec:galform_aligninternal}

\begin{table}
\begin{center}
\begin{tabular}{llll}
\hline
{median} & $z=0$ & $z=0.5$ & $z=1$ \\
\hline
$\cos\theta_{\rm S,G}$    & $0.993^{+0.001}_{-0.003}$ & $0.992^{+0.001}_{-0.003}$ & $0.991^{+0.001}_{-0.003}$ \\
$\cos\theta_{\rm S,DM}$ & $0.967^{+0.002}_{-0.013}$ & $0.932^{+0.006}_{-0.018}$ & $0.827^{+0.013}_{-0.072}$ \\
$\cos\theta_{\rm G,DM}$ & $0.949^{+0.004}_{-0.017}$ & $0.914^{+0.008}_{-0.025}$ & $0.817^{+0.016}_{-0.070}$ \\
\hline
$\cos\theta_{\rm S,tot}$    & $0.657^{+0.022}_{-0.062}$ & $0.690^{+0.027}_{-0.064}$ & $0.674^{+0.023}_{-0.068}$ \\
$\cos\theta_{\rm G,tot}$    & $0.652^{+0.023}_{-0.068}$ & $0.667^{+0.029}_{-0.058}$ & $0.711^{+0.020}_{-0.063}$  \\
$\cos\theta_{\rm DM,tot}$ & $0.712^{+0.023}_{-0.055}$ & $0.761^{+0.021}_{-0.072}$ & $0.827^{+0.013}_{-0.087}$ \\
\hline
\end{tabular}
\end{center}
\vspace{0.5cm}
\caption{\label{tab:GalaxyComponentAlignment} (Upper half) Median alignment angles between the gas (G),
stellar (S) and dark matter (DM) component of the galactic disk at redhifts $z=0$, $0.5$ and $1$. (Lower half)
Median alignment angles between the three components of the galactic disk and the total angular momentum
of the host halo. Errors are non-symmetric uncertainties on the median (see text for more details). Values are
given for the entire sample of well-resolved central galaxies since no mass dependence was found.}
\end{table}

\begin{figure}
  \begin{center}
    \includegraphics[width=0.3\textwidth]{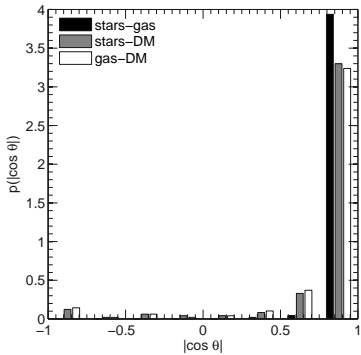}\\
    \vspace{-0.5cm}
  \end{center}
  \caption{ \label{fig:GalaxyComponentAlignment}Distribution of angles between the angular momentum 
  vectors of the stellar, gas and dark matter component within the blob that identifies the galaxy. The angle 
  distributions are shown for the entire sample of well-resolved central galaxies. }
\end{figure}

\subsection{Angular momentum correlations within the galaxy}
In Figure \ref{fig:GalaxyComponentAlignment}, we show the distributions of angles $\theta$ between the 
angular momentum vectors of the stellar (S), gas (G) and dark matter (DM) component within
well-resolved central galaxies at $z=0$ (the nomenclature is such that, e.g., $\theta_{\rm S,G}$ is 
the angle between the stellar and gas angular momentum vectors). The corresponding median angles are given 
for $z=0-1$ in Table \ref{tab:GalaxyComponentAlignment}
(upper half) together with the error on the median. We compute the upper error $\sigma_u$ 
and lower error $\sigma_l$ as
\begin{equation}
\sigma_u = \frac{\theta_{84}-\theta_{50}}{\sqrt{N}},\quad\sigma_l = \frac{\theta_{16}-\theta_{50}}{\sqrt{N}},
\end{equation}
where subscripted numbers indicate percentiles and $N$ is the number of galaxies considered.

Within the galactic disks (which is defined by the mesh cluster identified as a single galaxy -- see Section \ref{sec:galfind} -- 
and thus has no imposed symmetry for the stars and the gas), the gas angular momentum 
deviates by a median of 8 degrees from that of the stars. The difference with respect to the dark matter 
angular momentum, which was determined in a sphere around the disk, is slightly higher at around
18 degrees at $z=0$ for both stars and gas and increases to  around $36$ degrees at $z=1$.

We find no correlation between these alignment angles and either stellar or host halo mass
with a significance above $1\sigma$. This is a comforting result as stellar mass is directly
proportional to the number of star particles and the lack of a correlation demonstrates the
absence of an obvious resolution problem. We discuss numerical convergence of our
results in more detail in Appendix \ref{sec:numerical_artefacts}.

\subsection{Angular momentum correlations between central galaxies and their host halo}

\begin{figure*}
  \begin{center}
    \includegraphics[width=0.8\textwidth]{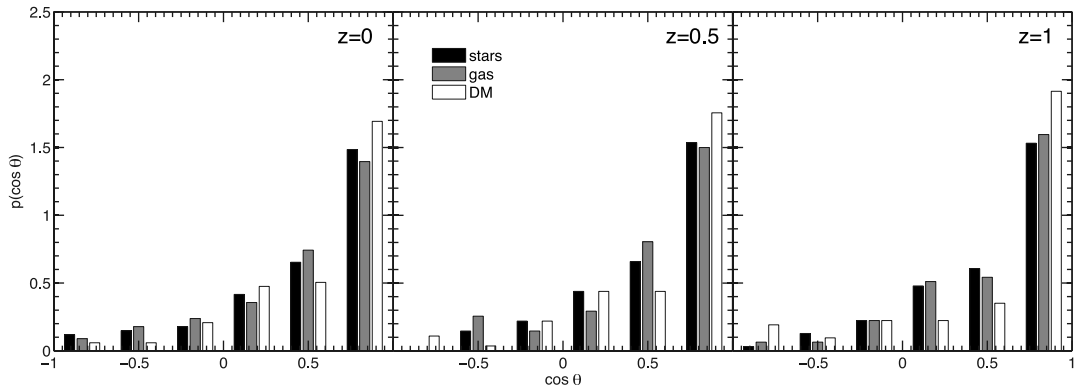}
  \end{center}
  \caption{ \label{fig:GalaxyHaloAlignment}Distribution of angles between the total (dark matter plus baryonic)
  angular momentum vectors of the dark matter host halo and the stellar, gas and 
  dark matter component of the central galaxy. The angle 
  distributions are shown for the entire sample of well-resolved central galaxies
  at $z=0$ (left), $z=0.5$ (centre) and $z=1$ (right panel). }
\end{figure*}

In Figure \ref{fig:GalaxyHaloAlignment} we show the distribution of angles between the 
angular momentum vector of the host dark matter halo (determined out to the virial radius
from both baryonic and dark matter)
and the inner components (determined at the disk) for the three redshifts at which we
perform our analysis. Galaxies from all samples have been combined. The corresponding median
angles and the width of the distributions are given in Table \ref{tab:GalaxyComponentAlignment}. 

We find a fairly broad distribution of angles with the host angular momentum deviating by
$\sim 50$ degrees at $z=0$ from either the stellar or gas spin of the central galaxy. 
Our results indicate that this discrepancy decreases slightly to about 46 degrees at $z=1$. 
The value we find for the stellar component is in excellent agreement with the results found by \cite{Croft2008} who
report a median angle for the stellar component of $44$ degrees and $70$ degrees
for the gas. The quoted larger angle for the gaseous component is possibly due to
satellites that are included in the sample of \cite{Croft2008} and for which ram pressure modifies the
gas disk orientations very quickly. 
It appears, however, that our results are in slight tension with the recent findings
of \cite{Bett2009} who report a smaller median angle of $34$ degrees between galaxy
and outer dark matter halo at $z=0$. It remains a matter of future studies to better quantify
the influence of resolution and feedback recipes on these results. A further possible origin for
the discrepancy between our results and those of \cite{Bett2009} might lie in the difference
of large scale environment. While we simulate the relatively high-pressure environment of
a massive filament, the galaxy sample considered by \cite{Bett2009} has a halo mass upper limit 
of $1.6 \times 10^{12} {\rm M}_\odot$, suggesting that they reside in a lower density
environment with a more quiescent formation history.

The outer and inner dark matter spins are
significantly stronger aligned than the baryonic components. We find a median angle of $\sim 45$ 
degrees at $z=0$ which decreases to $\sim 34$ degrees at $z=1$. 
Again, we find no correlation between the alignment angles and either stellar or host halo mass
with a significance above $1\sigma$.

\section{The Large-Scale Orientations of Galaxies}
\label{sec:galform_align_lss}
\subsection{\label{sec:analyticpred}Predictions of Analytic Models}
The linear tidal torque theory predicts that the angular momentum of a proto-galaxy
is determined by its geometric shape and the tidal field exerted by the larger scale
environment \citep[e.g.][]{Peebles1969,White1984,Porciani02}. In particular,
angular momentum is generated from the misalignment of the Lagrangian moment of 
inertia tensor $I_{ij}$ and the Lagrangian tidal tensor $T_{ij}$, 
so that at first order
\begin{equation}
J_i \propto \epsilon_{ijk}T_{jl}I_{lk},
\end{equation}
where $\epsilon_{ijk}$ is the Levi-Civita symbol. In the eigenspace of the 
tidal tensor, the components of ${\bf J}$ become
\begin{eqnarray}
J_1 & \propto & (\lambda_2-\lambda_3) I_{23},\nonumber\\ 
J_2 & \propto & (\lambda_3-\lambda_1) I_{31},\\
J_3 & \propto & (\lambda_1-\lambda_2) I_{12}\nonumber.
\end{eqnarray}
Since, by definition, $\lambda_3-\lambda_1$ is the largest coefficient, linear tidal torque 
theory thus predicts that, at first order, ${\bf J}$ is preferentially aligned with ${\bf v}_2$ in a 
statistical sample of haloes/galaxies where contributions from the moment of inertia
tensor average out and $T_{ij}$ and $I_{ij}$ are assumed to be uncorrelated. Moreover,
generation of angular momentum is most efficient at turnaround when the $I_{ij}$ is maximal.

However, the assumption that $T_{ij}$ and $I_{ij}$ are uncorrelated is not well justified, 
as the two tensors are indeed found to be correlated \citep[see e.g.][]{Porciani02, Lee2009b}. 
Thus it is particularly interesting to see whether the alignment predicted from Lagrangian 
perturbation theory is indeed present in hydrodynamic cosmological simulations, and particularly
so at late times well after turnaround of the proto-galaxy.

\cite{Lee2007} have derived a probability distribution for the angle  
$\cos \theta_2=|{\bf J}\cdot {\bf v}_2| / \|{\bf J}\|$ -- along which 
the possibly strongest alignment could be observed -- of the form
\begin{equation}
p(\cos \theta_2) = (1+c)\sqrt{1-\frac{c}{2}}\left[1+c\left(1-\frac{3}{2}\cos^2 \theta_2\right)\right]^{-3/2},
\end{equation}
where $c$ is a normalisation parameter.

\subsection{The Distribution of Spin and Tidal Field Eigenvectors}
\subsubsection{Anisotropy of the Tidal Field}
\begin{figure*}

  \begin{center}
  \includegraphics[width=0.25\textwidth]{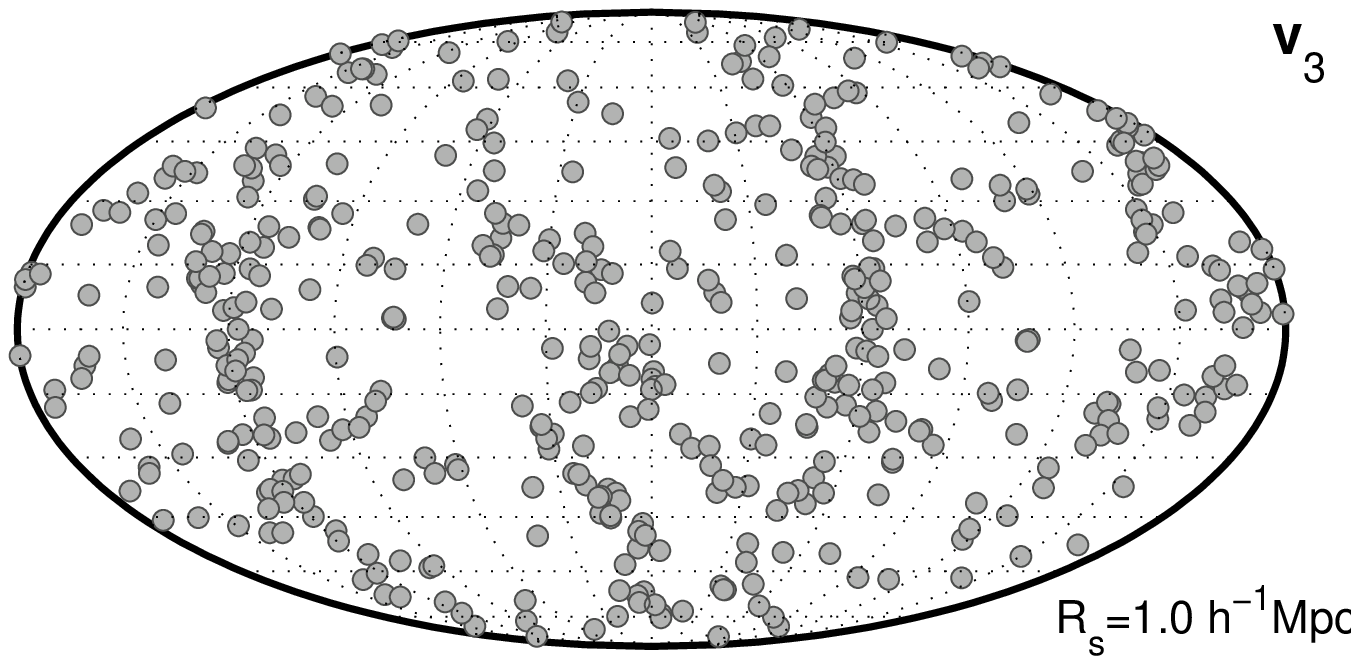}
  \includegraphics[width=0.25\textwidth]{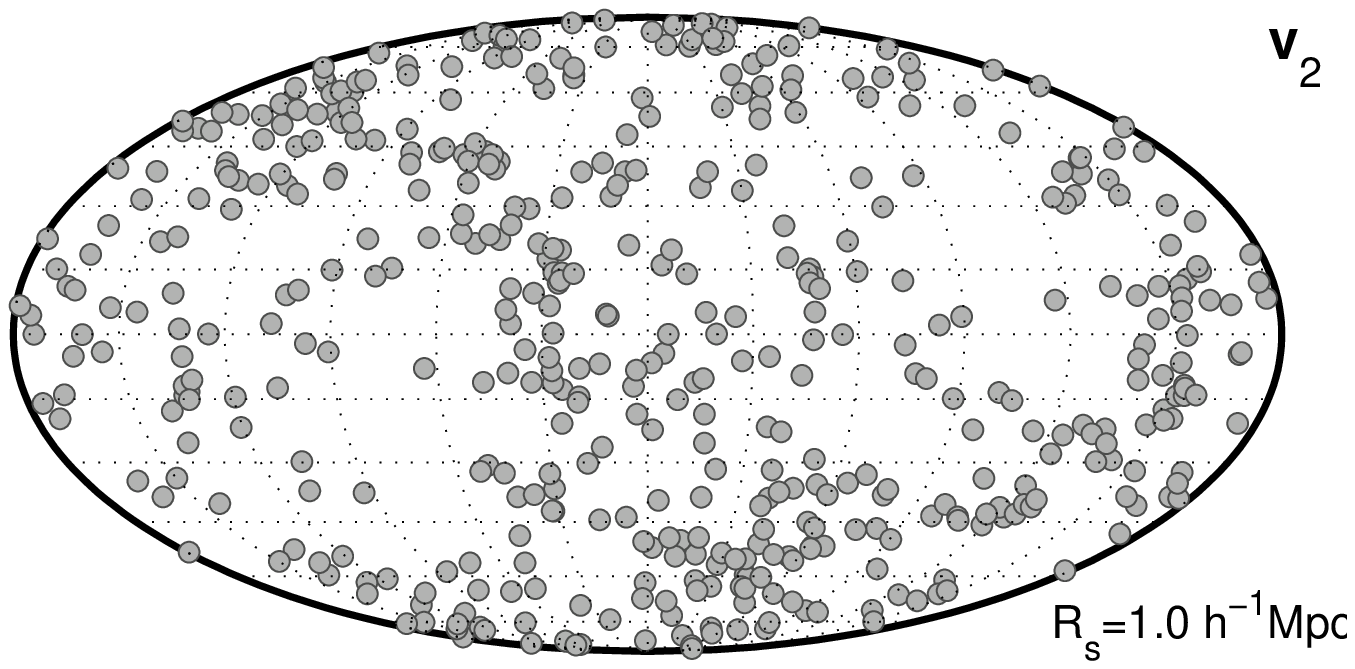}
  \includegraphics[width=0.25\textwidth]{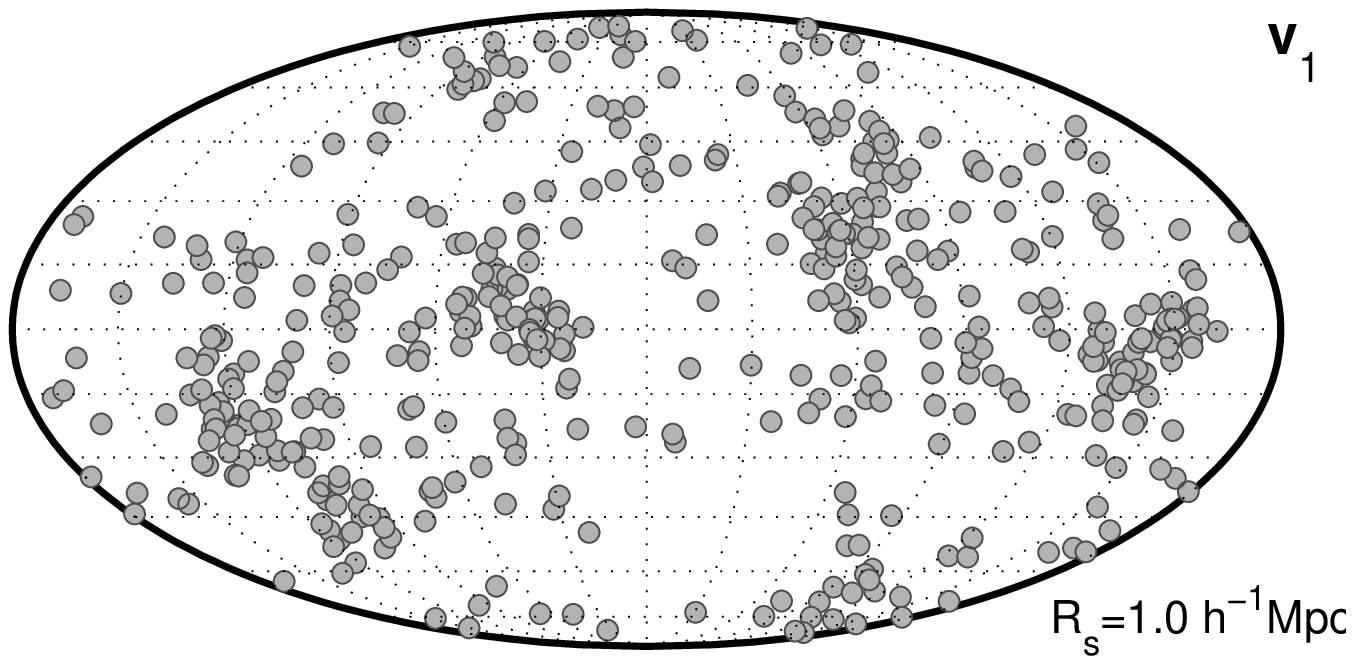}\\
  \includegraphics[width=0.25\textwidth]{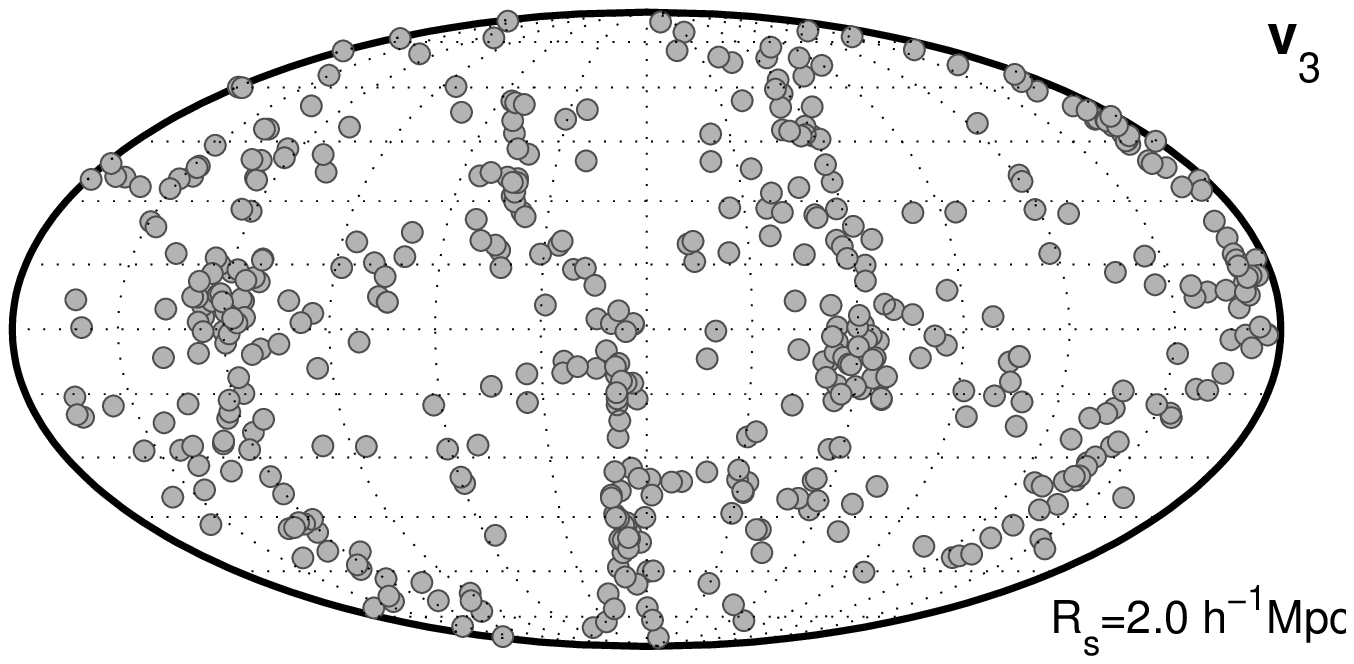}
  \includegraphics[width=0.25\textwidth]{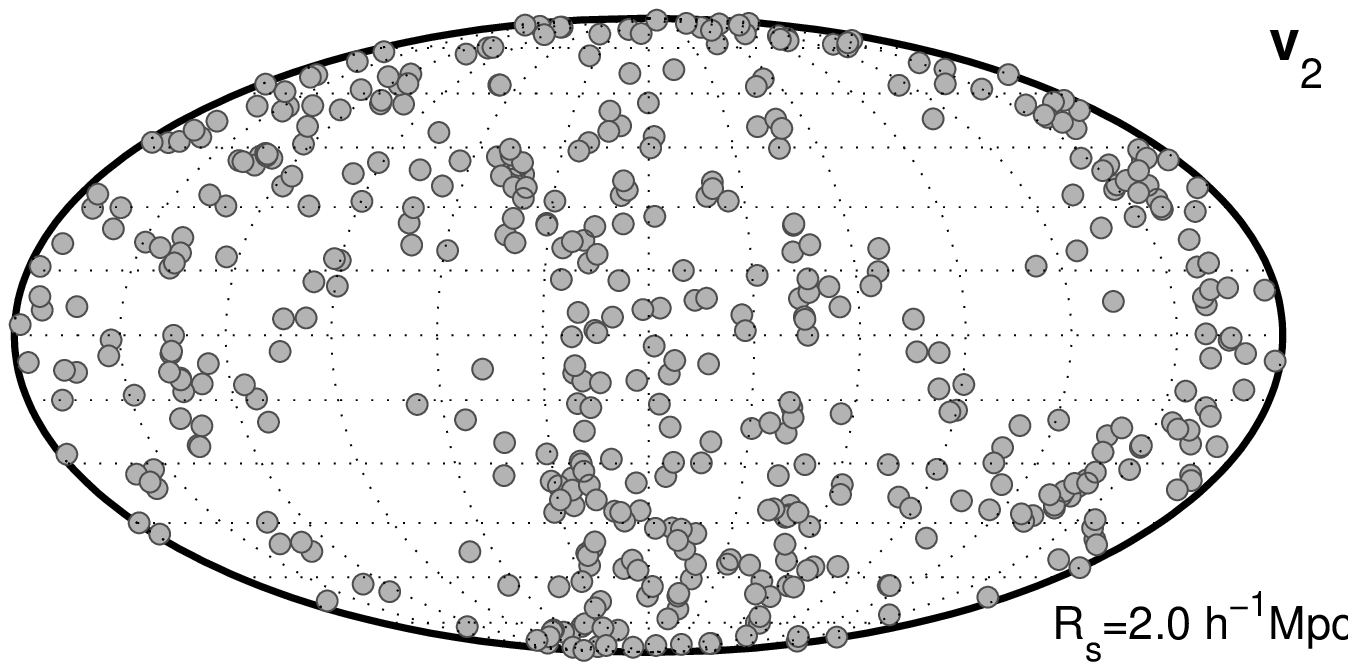}
  \includegraphics[width=0.25\textwidth]{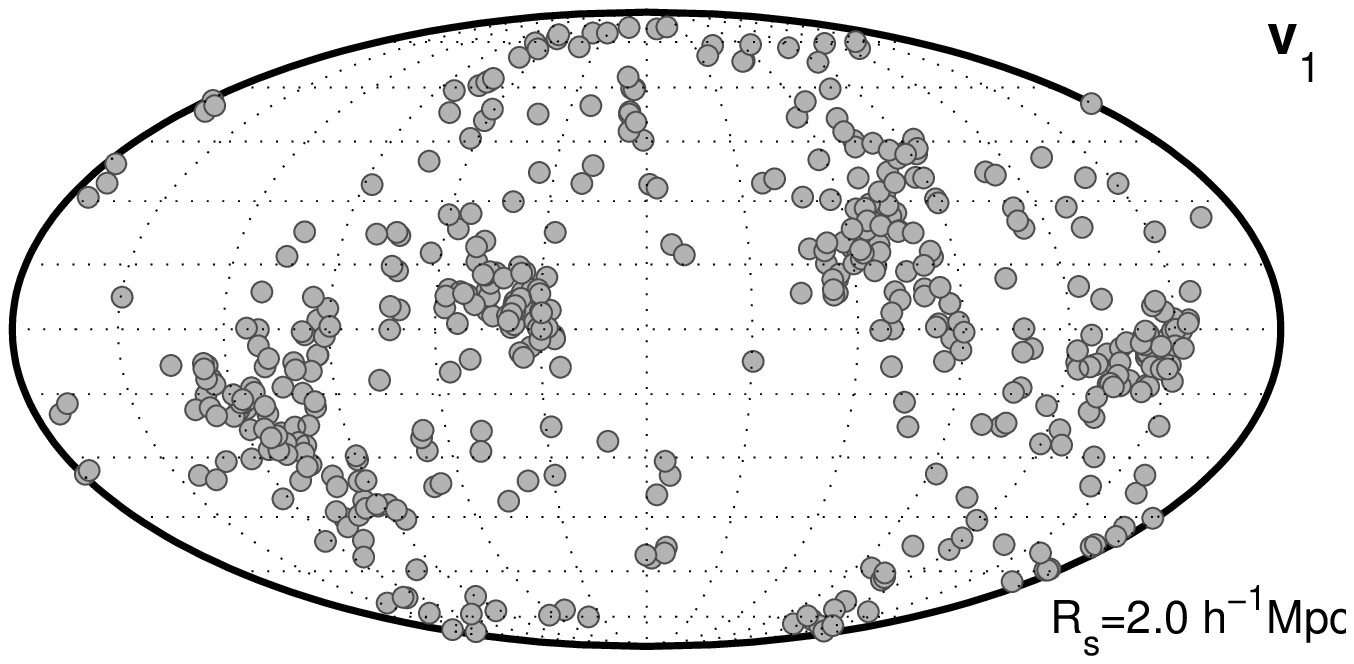}\\
  \includegraphics[width=0.25\textwidth]{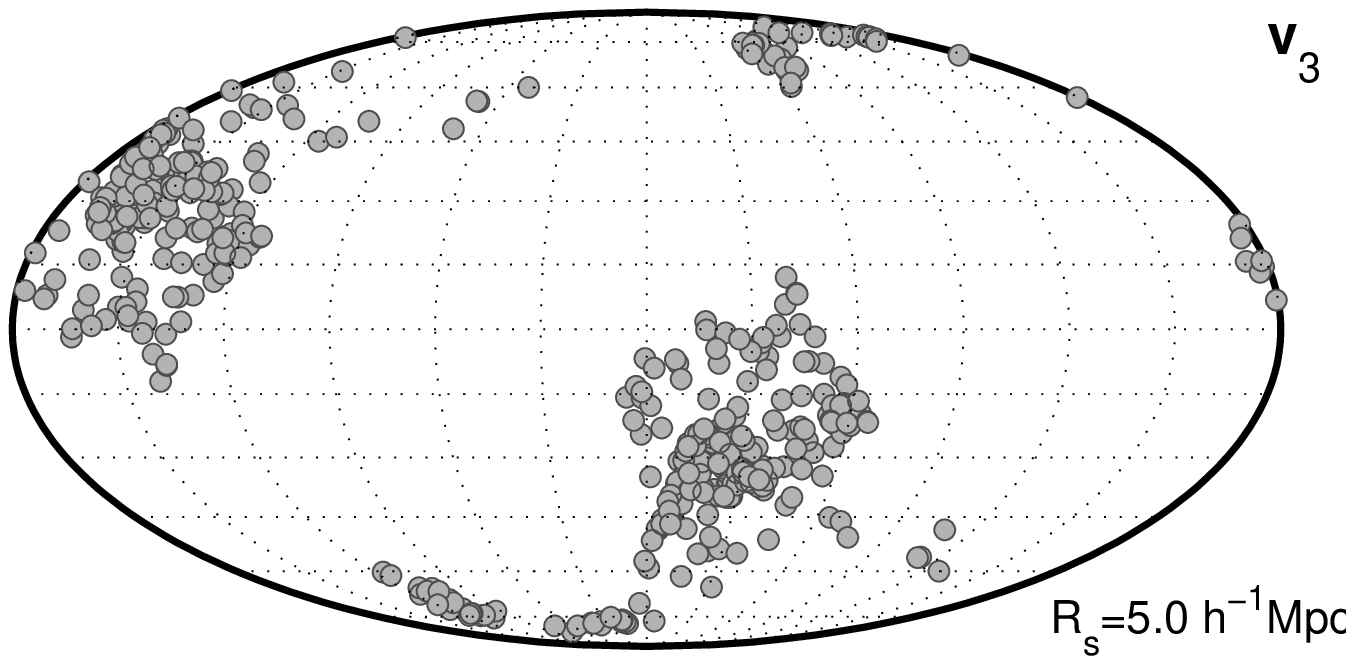}
  \includegraphics[width=0.25\textwidth]{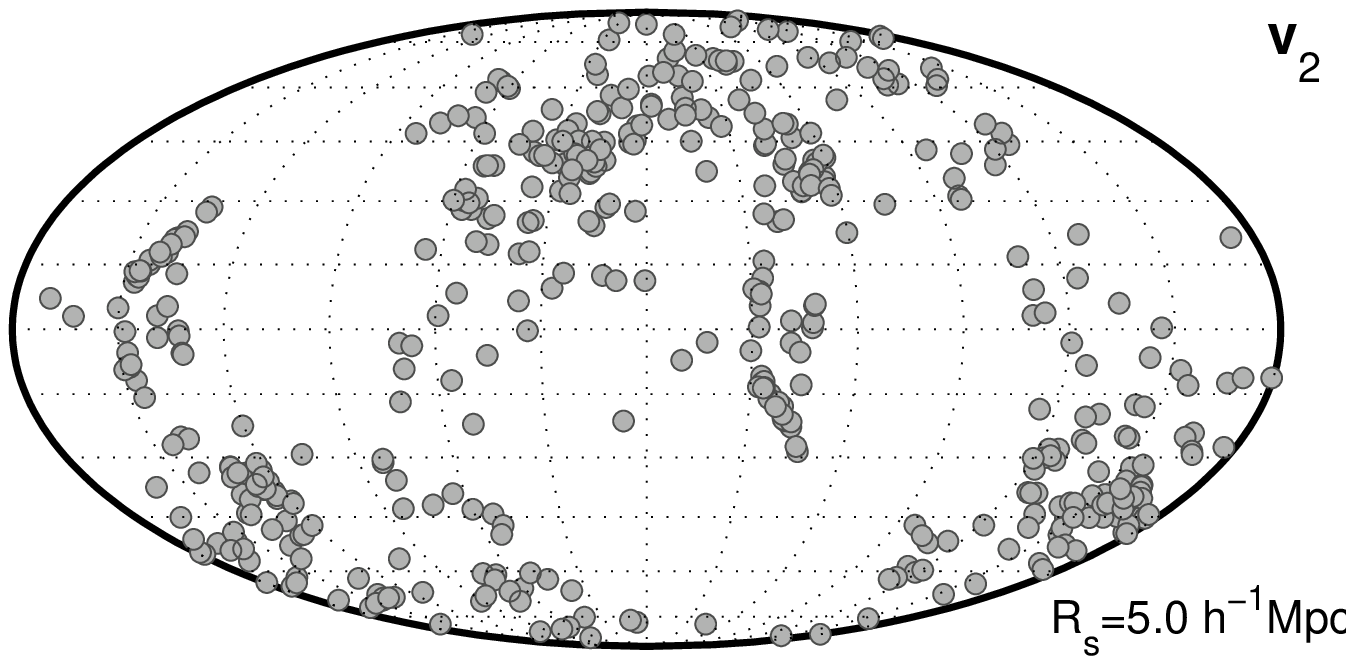}
  \includegraphics[width=0.25\textwidth]{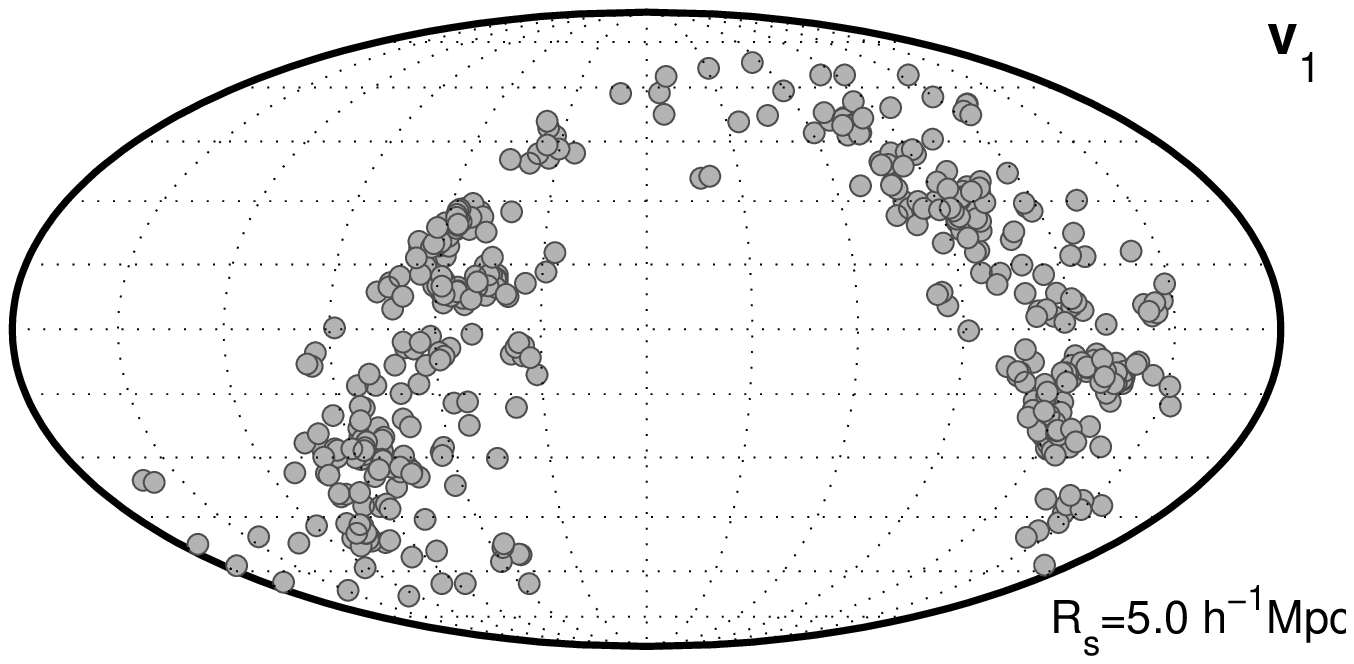}\\

  \end{center}
  
  \caption{\label{fig:ProjTides} Mollweide sphere projections of the three eigenvalues of the 
  tidal field at the positions of the galaxies for three smoothing scales, $1\,h^{-1}{\rm Mpc}$ (top),
  $2\,h^{-1}{\rm Mpc}$ (middle) and $5\,h^{-1}{\rm Mpc}$ (bottom). All data is
  for redshift $z=0.5$. Since the tidal field has a spin-2 symmetry, each vector is represented
  by two points. The Mollweide projection is an area preserving projection.}
\end{figure*}

In Figure \ref{fig:ProjTides}, we show the spatial distribution at $z=0.5$ of tidal field eigenvectors ${\bf v}_{1\dots3}$
for the comoving smoothing scales $R_s=1$, 2 and $5\,h^{-1}{\rm Mpc}$ 
in Mollweide projection. The tidal field is evaluated at the positions of the galaxies that are included
in any of the samples (LM or HM). As this is an equal-area projection, local concentrations directly
 translate into an increased anisotropy for that direction. Since the tidal field is a tensor field,
each vector has spin-2 symmetry and hence is represented by two points, the second being
a reflection of the first at the origin. While the tidal field is already highly anisotropic on scales of
$1\,h^{-1}{\rm Mpc}$, on scales of $5\,h^{-1}{\rm Mpc}$ virtually only the main filament remains
so that the distribution of eigenvectors becomes almost dipolar.

\subsubsection{Anisotropy of the Galaxy Spin Distribution}
\begin{figure*}
	\begin{center}
	  \includegraphics[width=0.3\textwidth]{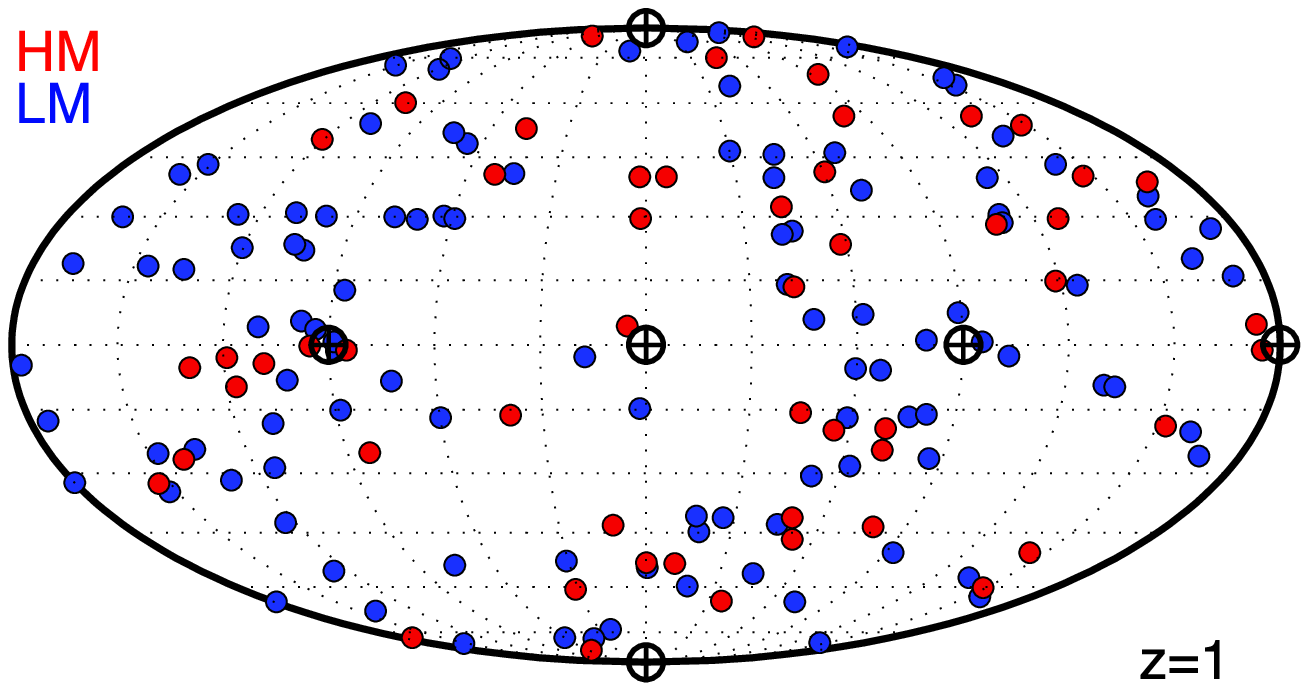}
	  \includegraphics[width=0.3\textwidth]{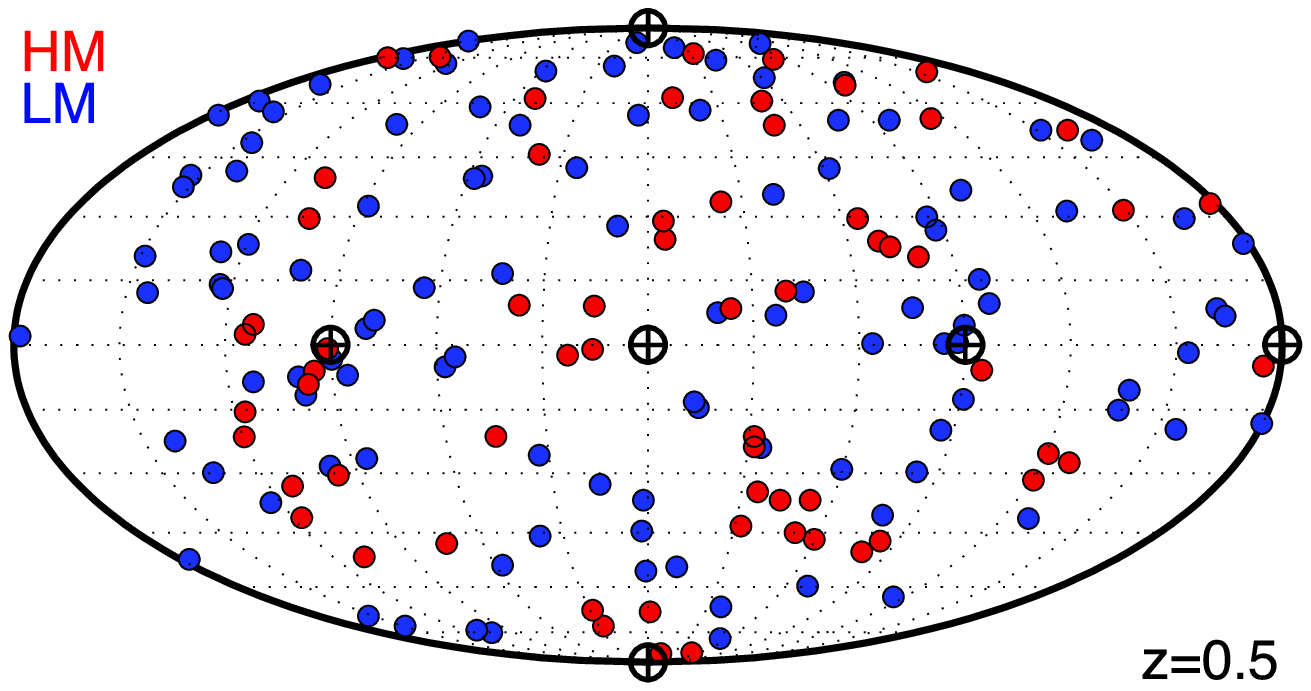}
	  \includegraphics[width=0.3\textwidth]{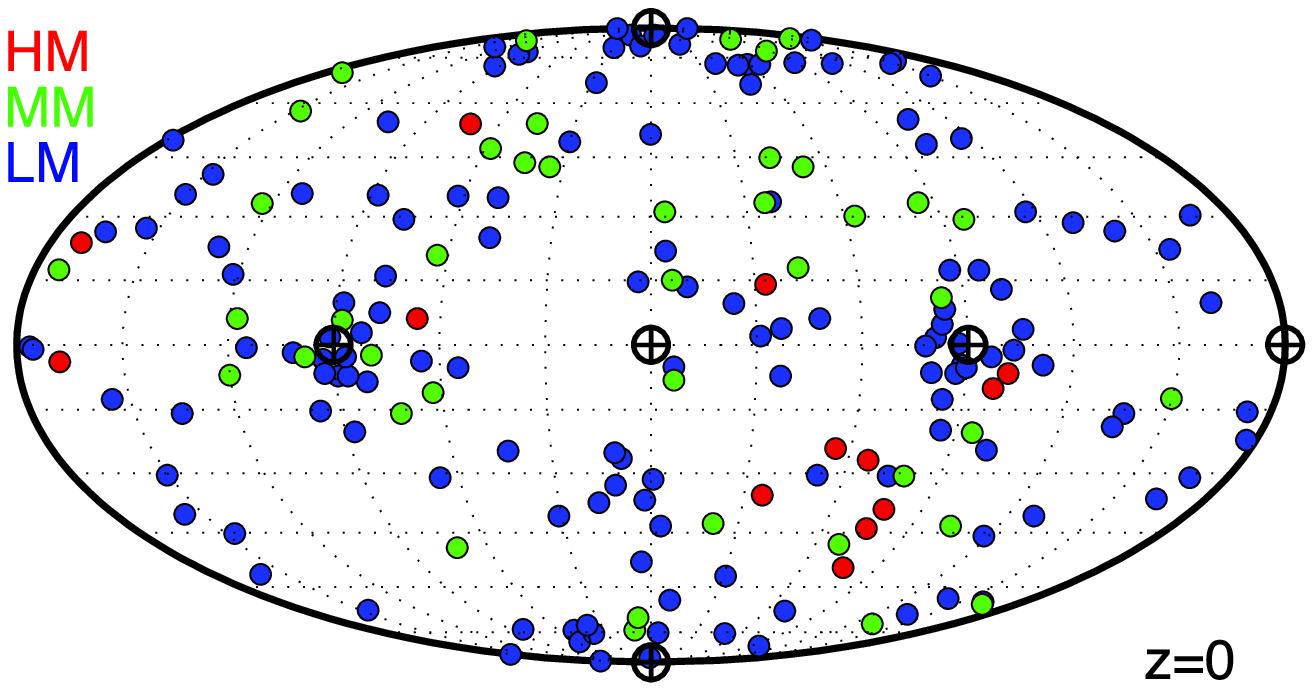}
	\end{center}
	\caption{\label{fig:ProjSpins}The orientations of galaxy stellar angular momenta at redshifts $z=1$ (left),
	$z=0.5$ (middle) and $z=0$ (right). The angular momentum vectors are shown in the Mollweide projection.
	The circled crosses indicate the Cartesian basis vectors of the AMR grid.}
\end{figure*}

Figure \ref{fig:ProjSpins} shows the distribution of stellar angular momentum vectors at redshifts
$z=1$, 0.5 and 0 in Mollweide projection. At all redshifts the distributions are clearly anisotropic.
The origin and detailed form of this anisotropic distribution is determined by both physical and 
numerical effects. We will discuss their interaction in what follows.

At $z=1$, the distribution is anisotropic with no clear clustering of spin vectors around the Cartesian
grid vectors. This provides strong evidence that the high-redshift spin distribution is anisotropic due
to physical processes. Comparing with Figure 6, the tidal field on scales $<5\,h^{-1}{\rm Mpc}$ is
anisotropic at a similar level. We will discuss their cross-correlation in the next section.

At $z=0.5$ there is a slight tendency for the spins to cluster around the grid vectors that is more obvious
at $z=0$ for the low-mass galaxies. Since a multi-grid Poisson solver (such as the one used in {\sc Ramses})
introduces preferred directions along the Cartesian grid
basis vectors, a non-physical anisotropy of the gravitational inter-particle force is expected close to the resolution limit
\citep[cf.][]{Hockney1981} arising from an unavoidable lack of symmetry in the discrete Laplacian operator. 
These errors are systematic in any asymmetric system \citep[cf.][]{May1984}.
This is clearly a systematic effect in our numerical results, and we need to estimate its magnitude.

Remarkably, the grid-aligned galaxies at low redshift are not distributed isotropically among all Cartesian
grid vectors. In larger scale simulations of entire cosmological volumes (i.e. not restricted to a filamentary region --
as e.g. in the ``Mare Nostrum'' simulation, \citealp{Ocvirk2008}), we do not see this effect. In these larger volume simulations
at lower spatial resolution, grid-alignment is also observed but with galaxies aligned with all three grid vectors.

Given the highly anisotropic distribution of tidal field
eigenvectors (see previous subsection), it is however clear how such an anisotropy can arise. Assuming that
the galaxies' orientations are determined by some anisotropic physical effect (such as tides), the disks' 
initial angular momentum will be anisotropic (as it is at $z=1$). Given the additional numerical anisotropy, the disks will
subsequently relax slowly to the closest potential minimum, i.e. the closest Cartesian grid vector, in the absence
of any perturbation that is larger than the numerical error (and thus predominantly in low density regions). 
This will happen on a timescale that is connected to the 
anisotropy of the error in the multi-grid Poisson solver. Support for this interpretation of the numerical results
comes from a look at the galaxies' spin distribution at high redshift where no obvious grid-alignment is 
observed. Looking at the left panel of Figure \ref{fig:ProjSpins}, we see a non-isotropic distribution with no
spins in the vicinity of those grid axes where by $z=0$ also no grid-aligned galaxies are seen (e.g. the one
in the very centre of the Mollweide maps).

We will continue to discuss the impact of such a spurious alignment on our results in Section \ref{sec:numerical_artefacts}.
To conclude this section, we can however clearly say that the distribution of galaxy angular momenta is anisotropic
due to galaxy formation physics. The anisotropy is then likely amplified by numerical effects at low redshifts, but numerical effects
alone will never generate an anisotropic distribution.

\subsection{\label{sec:corr_tides}Correlations with the tidal field}

\begin{figure*}
  \begin{center}
  \includegraphics[width=0.7\textwidth]{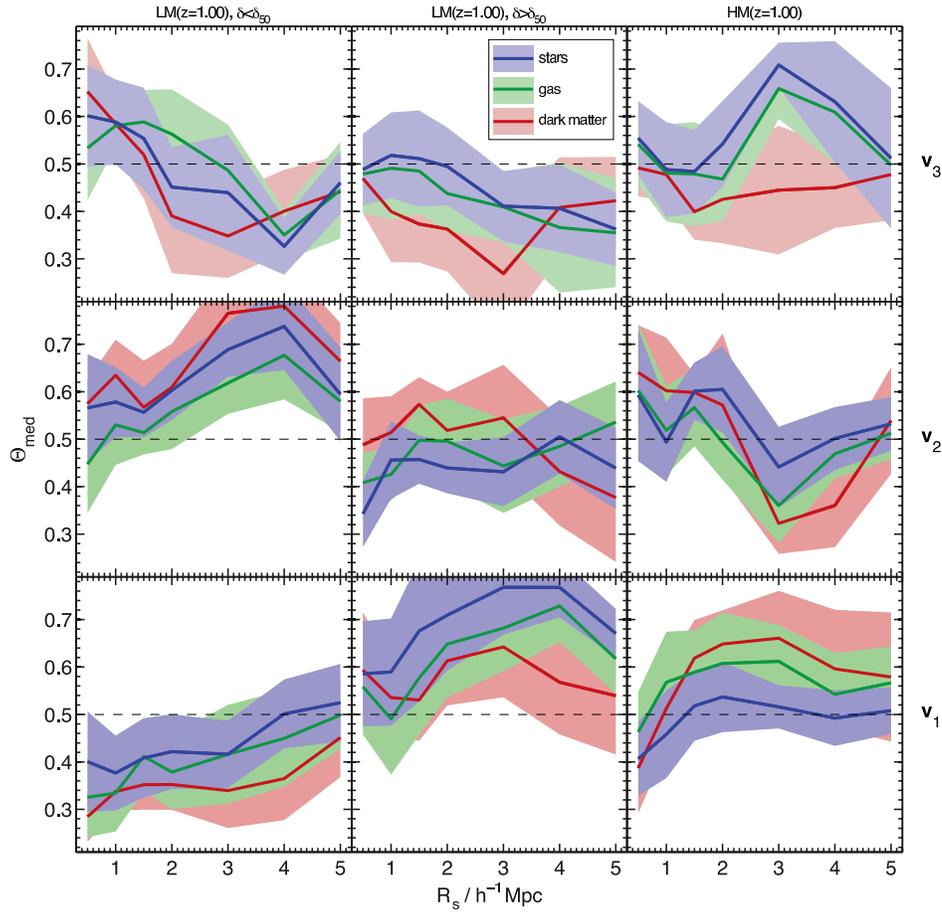}
  \end{center}
  \caption{\label{fig:SpinTide00145} Median cosine at $z=1$ of the angle between the spin of the
  dark matter (red), gaseous (green) and stellar (blue) component of central galaxies and
  the three eigenvectors of the tidal field tensor as a function of the scale $R_s$ (given in
  $h^{-1}\,{\rm Mpc}$ comoving) on which the
  tidal field was smoothed. The data is shown in for the halo mass bins
  given in Table \ref{tab:halo_samples} with the low mass (LM) sample split into two samples
  above and below the median environmental matter overdensity determined on physical scales
  of $2\,h^{-1}{\rm Mpc}$ (i.e. $4\,h^{-1}{\rm Mpc}$ comoving). The shaded areas represent the $1\sigma$
  uncertainty on the median obtained with 200 bootstrap resamplings.}
\end{figure*}

As outlined in Section \ref{sec:analyticpred}, linear tidal torque theory indicates that galaxies
might show a preferential alignment with the intermediate principal axis of the tidal field. In this
Section, we investigate whether such an alignment persists also at times after the turnaround of
the proto-galaxy in our simulation.  Unfortunately, the scale on which the tidal field should be 
computed is not known a priori so that we present the degree of alignment for a range of smoothing
scales. Since these scales are likely to be different for galaxies of different masses, and galaxies
of different halo masses may also experiences differences in the details of their formation, all results 
are split into different halo masses as described in Section {sec:galsamples}. In addition, we 
distinguish low-mass galaxies in high-density and in low-density environments.

In Figure \ref{fig:SpinTide00145}, we show the scale dependence of the cosine of the angles $|\cos\theta_i|$
between the angular momentum vectors of the galaxies and the tidal field eigenvectors
${\bf v}_i$ for the inner dark matter, gas and stellar components. The panels represent
the alignments with the 3 eigenvectors (vertical) for the 2 mass bins (horizontal). We
additionally split the low mass bin (LM) into an overdense bin $\delta>\delta_{50}$ and an 
underdense bin $\delta<\delta_{50}$, where here $\delta$ is the matter overdensity field smoothed on 
a $2\,h^{-1}{\rm Mpc}$ physical scale and $\delta_{50}$ is the median overdensity for the 
galaxies in the respective mass bin. The values of $\delta_{50}$ for our sample of galaxies
turn out to be: 0.27 at $z=1$, 0.54 at $z=0.5$ and 1.16 at $z=0$.

\begin{table*}
\begin{center}
\begin{tabular}{ll|lll|lll|llll}
\hline
significance in &$\quad$& z=1 & & $\quad$ & z=0.5  & & $\quad$ & z=0 & &  & \\
units of $\sigma$ && ${\rm LM}$ & LM & HM & LM & LM & HM & LM & LM & MM & HM  \\
& &  $\delta<\delta_{50}$ & $\delta>\delta_{50}$ & & $\delta<\delta_{50}$ & $\delta>\delta_{50}$ & & $\delta<\delta_{50}$ & $\delta>\delta_{50}$ & & \\
\hline
$\theta_{3,{\rm G}}$ && 0.60 & 1.96 & 0.22 & {\bf 2.26} & 0.02 & 0.18 & 0.29 & 0.24 & 0.46 & 0.80\\
$\theta_{2,{\rm G}}$ && {\bf 2.30} & 1.07 & 0.26 & {\bf 2.27} & 0.15 & 0.52 & 0.01 & 0.92 & 1.06 & 0.45\\
$\theta_{1,{\rm G}}$ && 1.93 & 1.20 & 0.55 & 0.94 & 0.34 & 1.06 & 0.49 & 0.11 & 0.32 & 1.53  \\
\hline
$\theta_{3,{\rm S}}$ && 1.14 & {\bf 2.97} & 0.55 & {\bf 2.58} & 0.23 & 0.13 & 0.40 & 0.05 & 0.02 & 0.59 \\
$\theta_{2,{\rm S}}$ && {\bf 2.75} & 1.28 & 0.20 & 1.66 & 0.30 & 0.29 & 0.73 & 0.29 & 1.21 & 0.41 \\
$\theta_{1,{\rm S}}$ && {\bf 2.21} & 0.67 & 1.06 & 0.38 & 0.22 & 0.51 & 1.00 & 0.26 & 1.01 & 1.61 \\
\hline
$\theta_{3,{\rm DM}}$ && 1.28 & 0.32 & 0.69 & 1.98 & 1.46 & 0.68 & 0.01 & 0.72 & 0.36 & 0.92\\ 
$\theta_{2,{\rm DM}}$ && {\bf 3.43} & 1.05 & 1.21 & 1.56 & 0.11 & 0.94 & 1.72 & 0.40 & 0.38 & 0.79 \\
$\theta_{1,{\rm DM}}$ && 1.28 & 1.45 & 0.58 & 0.19 & 0.39 & 0.25 & 1.75 & 0.24 & 0.94 & 1.05 \\
\hline
\end{tabular}

\end{center}
\vspace{0.5cm}
\caption{\label{tab:sig_levels}Significance levels for the rejection of the null-hypothesis that the angle distributions 
between the spin of the gas disk (G), stellar disk (S) or inner dark matter (DM) with the tidal field eigenvectors are 
consistent with a uniform distribution $p(|\cos \theta_i|)=1$ obtained with the Kolmogorov-Smirnov test. The tidal
field was smoothed on a scale of physical $2\,h^{-1}{\rm Mpc}$ for this test. Boldface numbers highlight those
correlations that have a significance above $2\sigma$.}
\end{table*}

Using a two-tailed Kolmogorov-Smirnov test, we assess the inconsistency of the angle distributions
on a scale of physical $2\,h^{-1}{\rm Mpc}$ with a flat (i.e. random distribution) $p_{\rm rand}(|\cos\theta_i|)=1$.
The significance levels are given in units of $\sigma$ in Table \ref{tab:sig_levels}.

At $z=1$, we find the strongest alignment for the low-mass low-density sample, where on scales of 
physical $2\,h^{-1}{\rm Mpc}$, the alignment signal with the intermediate principal axes of the
tidal field peaks at a median angle of $|\cos\theta_2|\sim 0.7$
for the stellar angular momentum component. The deviation from a random angle distribution is significant
at $2.75\sigma$. The alignment of the gas disk is slightly weaker, while that of the inner dark matter
is slightly stronger. There is a weaker counter-alignment with the third principal axis on these scales,
while the alignment with the first axis is consistent with random.

For the low-mass high-density sample, we find a significant alignment only with the first
principal axis for the stellar and gas component, while the dark matter angular momentum
is consistent with random at almost all scales.

Finally, for the high-mass sample, we find a significant alignment of the stellar and gas component
with the third principal axis on comoving scales of $3\,h^{-1}{\rm Mpc}$ which is however not 
seen for the inner dark matter component at $z=1$. At lower redshifts, this alignment decreases
slightly but is also seen for the dark matter component.


\begin{figure*}
  \begin{center}
  \includegraphics[width=0.7\textwidth]{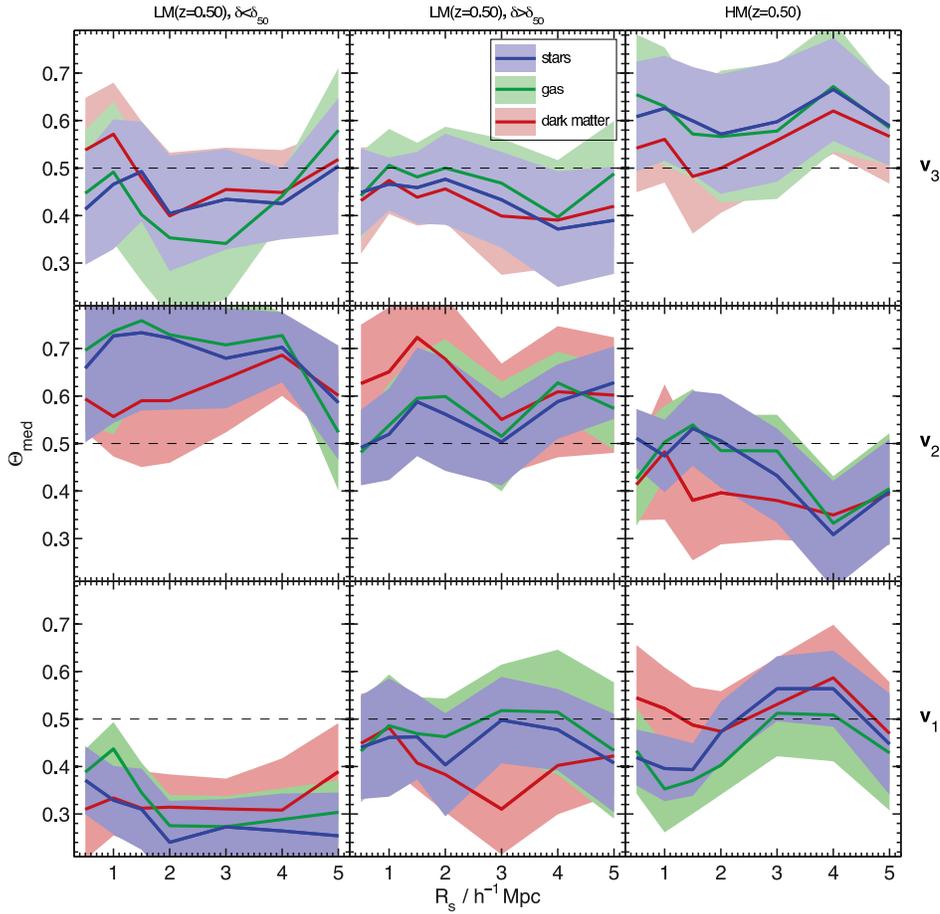}
  \end{center}
  \caption{\label{fig:SpinTide00257}Same as Figure \ref{fig:SpinTide00145} but at 
  redshift $z=0.5$. }
\end{figure*}
In Figure \ref{fig:SpinTide00257}, we show the scale dependent median alignment for the $z=0.5$
samples of galaxies. As for the $z=1$ sample, we again find a preferential alignment of the
low-mass low-density sample with the intermediate axis of the tidal field peaking on scales
of physical $2\,h^{-1}{\rm Mpc}$ and a counter-alignment with the first principal axis.
For the low-mass high-density environment, our findings are consistent with random.
There is again a weak tendency for the high mass sample to be aligned with the third
principal axis.


\begin{figure*}
  \begin{center}
  \includegraphics[width=0.9\textwidth]{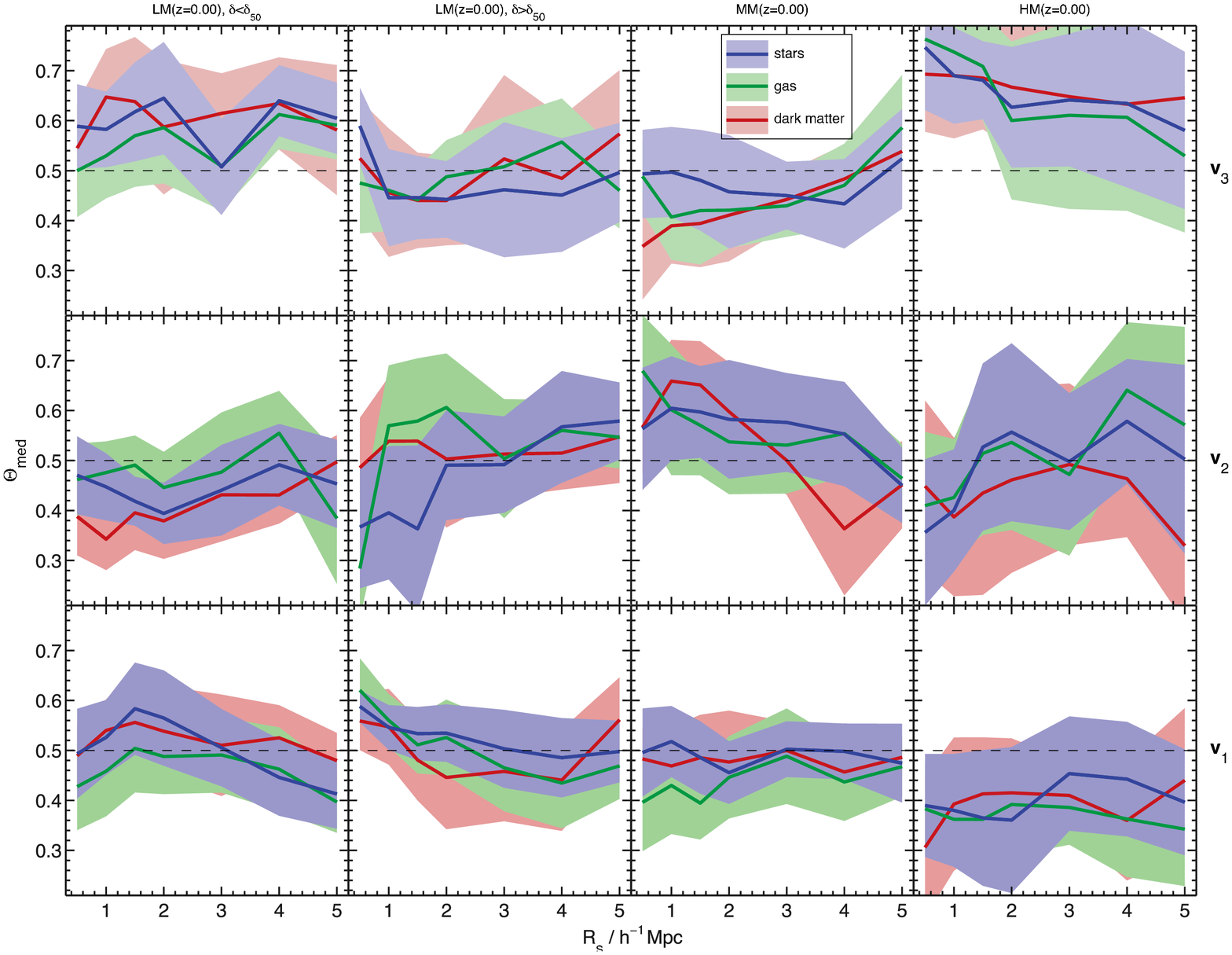}
  \end{center}
  \caption{\label{fig:SpinTide00439}Same as Figure \ref{fig:SpinTide00145} but at 
  redshift $z=0$. }
\end{figure*}
In Figure \ref{fig:SpinTide00439}, we show the scale dependent median alignment for the $z=0$
samples of galaxies. For the low-mass samples, all alignment signals are consistent with random.
For the medium mass sample, we find a very weak tendency for alignment with the intermediate
principal axis of the tidal field. The high-mass sample shows a significant alignment with the
third principal axis on small scales $\lesssim 1\,h^{-1}{\rm Mpc}$ that is still present but weakens 
at larger scales. Note however, that the high-mass sample consists only of 12 galaxies.

\begin{figure*}
  \begin{center}
  \includegraphics[width=0.6\textwidth]{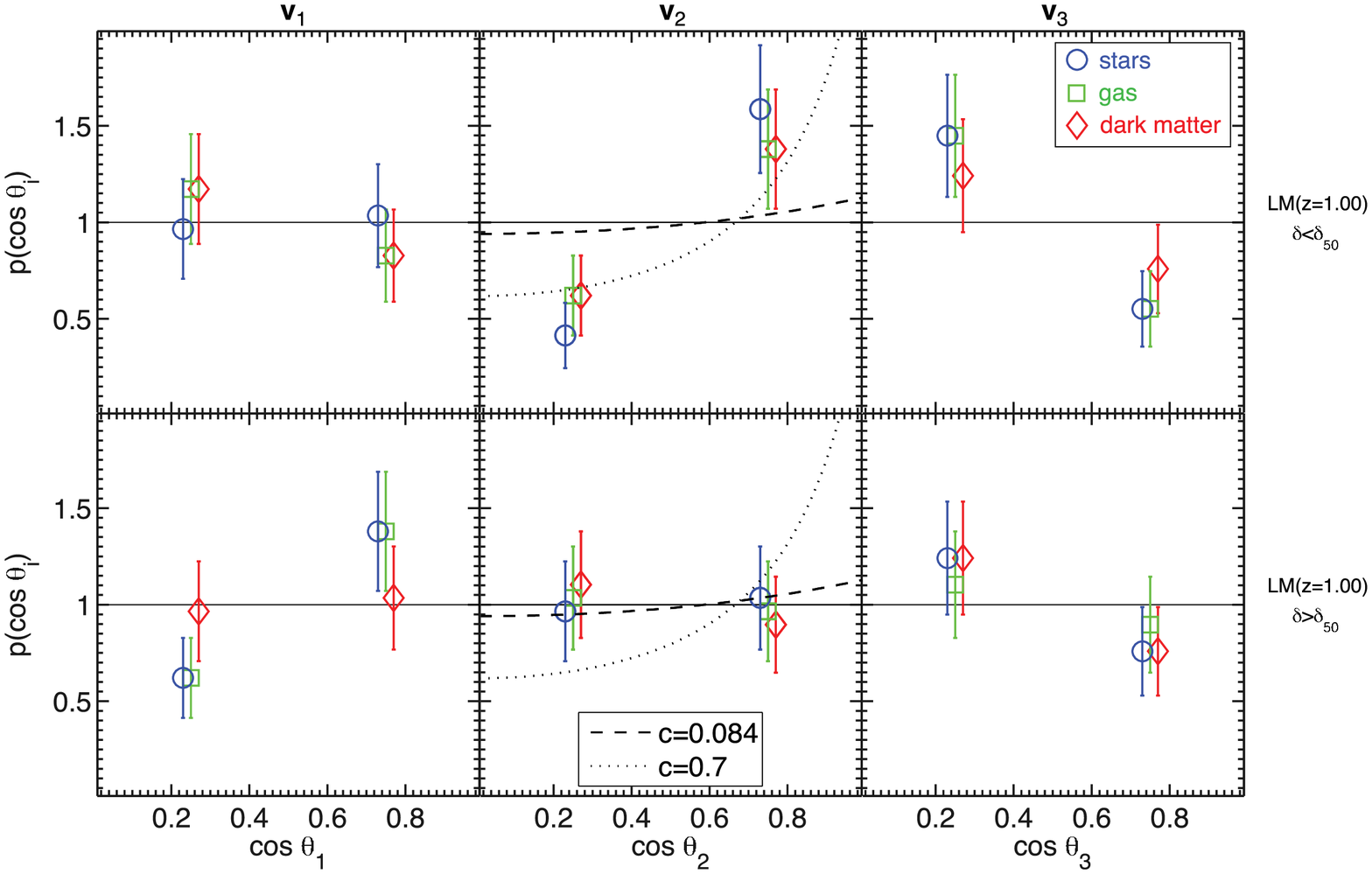}
  \end{center}
 \caption{\label{fig:PCosTide00145} The distribution of $|\cos \theta_i|$ between the galaxy spin vector
  and the three eigenvectors ${\bf v}_3$ (left), ${\bf v}_2$ (middle) and ${\bf v}_1$ (right) at $z=1$. 
  Shown is the binned median, errorbars correspond to the Poisson error on the median. The distributions are shown 
  for the stellar (blue circles), gaseous (green squares) and dark matter (red diamonds) components of the galaxies. The dashed and
  dotted line indicate the model predictions from \citet{Lee2007}.The dashed line corresponds to the observationally determined
  fit of the model as given in \citet{Lee2007}, while the dotted line corresponds to the model with $c=0.7$. In each plot, 
  the top panel shows the distribution for the low-mass low-density (LM, $\delta<\delta_{50}$) sample, and the lower the 
  distribution for the low-mass high-density (LM, $\delta>\delta_{50}$) sample. Density and alignments have been
  determined on scales of $2\,h^{-1}{\rm Mpc}$ physical.}
\end{figure*}

\begin{figure*}
  \begin{center}
  \includegraphics[width=0.6\textwidth]{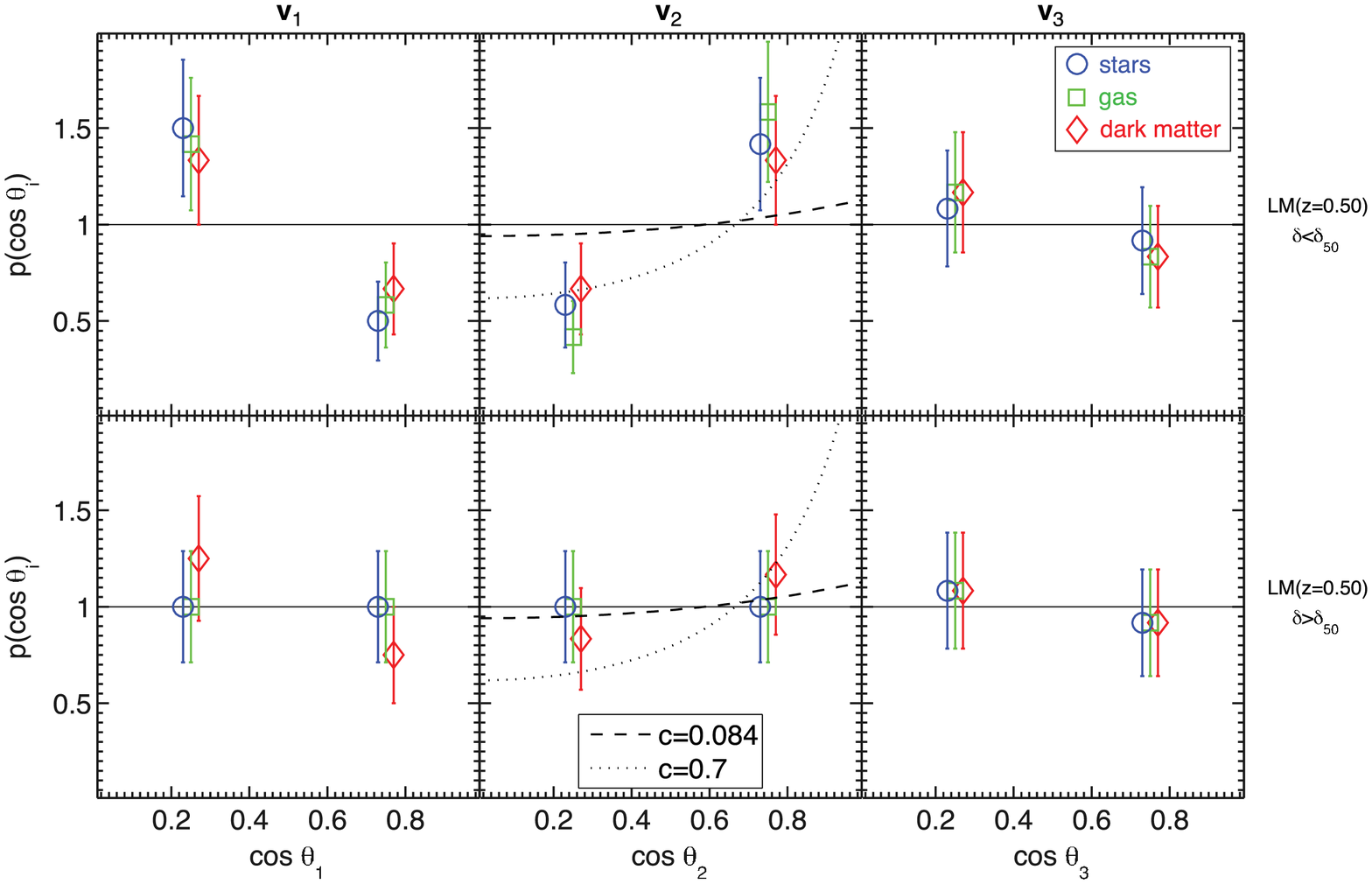}
  \end{center}
  \caption{\label{fig:PCosTide00257}Same as Figure \ref{fig:PCosTide00145} but for the low mass (LM) samples at $z=0.5$,
  split in low (top) and high (bottom) environmental density.}
\end{figure*}

\begin{figure*}
  \begin{center}
  \includegraphics[width=0.6\textwidth]{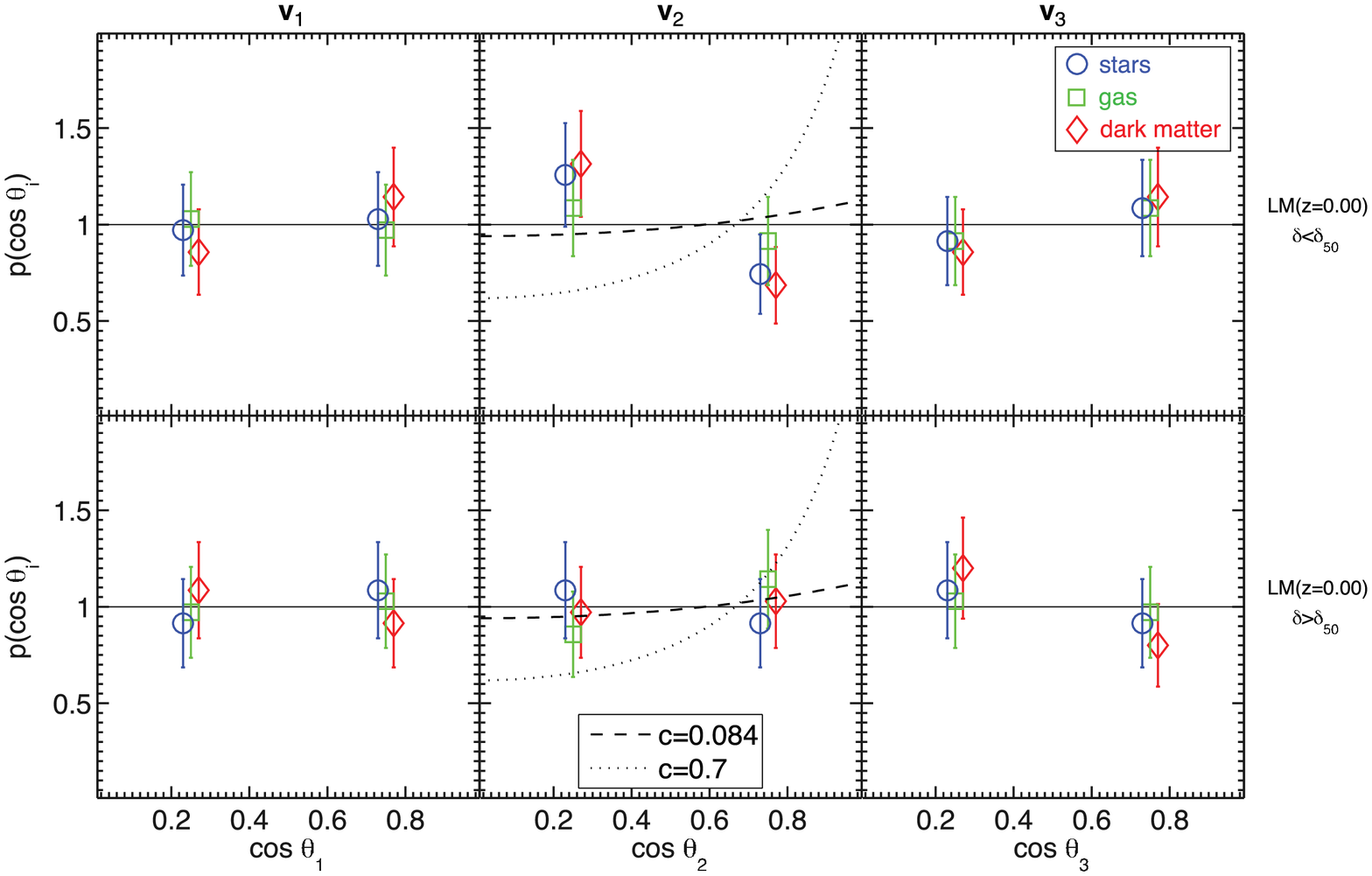}
  \end{center}
  \caption{\label{fig:PCosTide00439} Same as Figure \ref{fig:PCosTide00145} but for the low mass (LM) samples at $z=0$,
  split in low (top) and high (bottom) environmental density.
  }
\end{figure*}

\subsection{Alignment Angle Probability Distributions}
Figures \ref{fig:PCosTide00145}, \ref{fig:PCosTide00257}, and \ref{fig:PCosTide00439} show the probability
distributions $p(|\cos\theta_i|)$ for the low mass samples (LM) at redshifts $z=1$, 0.5 and 0, respectively. 
The galaxy samples have again been split into a high and low environment density sample above and below
the median density, as described in Section \ref{sec:corr_tides}, and the angle distributions are given
for the same scale on which the density field has been obtained -- $2\,h^{-1}{\rm Mpc}$ physical
at all redshifts.
 
For the higher redshifts $z=1$ and $0.5$, the low density sample is consistent with the model
fit for spin alignment with the intermediate axis of the tidal field given by \citet{Lee2007}
 albeit for a parameter which is significantly larger (roughly
a factor of 9) than the best fit value to the 2MASS galaxy samples obtained in their study.  
While this discrepancy might seem large, it is not clear to what extent our samples are 
comparable at all with the observational data from the 2MASS survey. Furthermore,
the weaker alignment seen in observational data might be due to the inherent difficulty to 
infer the orientation of a galaxy from the projected image, as well as uncertainties
in determining the tidal field. 

More curiously however, \citet{Lee2007} find a stronger alignment signal with the intermediate
axis for galaxies in high density regions. At all redshifts, for the high-density sample of 
low-mass galaxies, our results are consistent with random orientations. While tidal torque
theory predicts that alignment with the intermediate axis should indeed be stronger in
high density regions \citep{Lee2007}, our results do not support that this prediction  is valid 
in highly non-linear environments.

\subsection{The Impact of Environment on Alignments}
\begin{figure*}
  \begin{center}
  \includegraphics[width=0.7\textwidth]{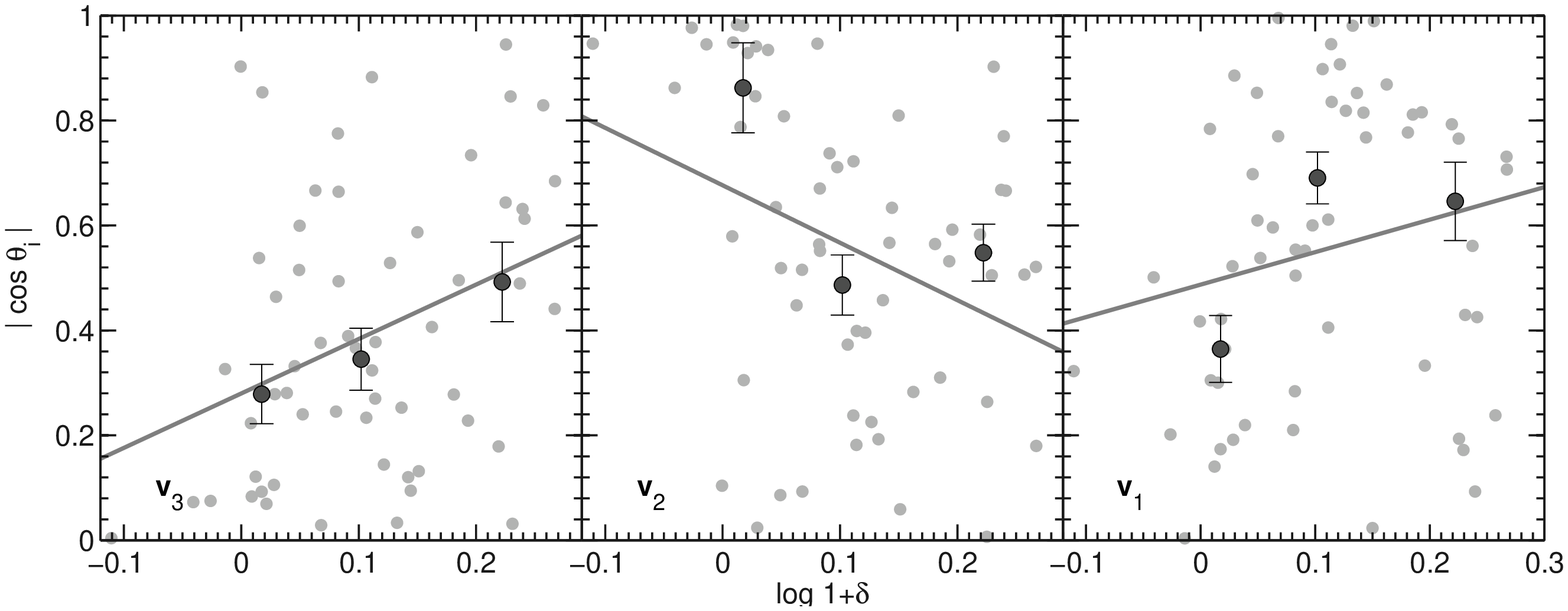}
  \end{center}
  \caption{\label{fig:delta00257}Cosine of the alignment angle of the stellar angular momentum
  with the three principal axes of the tidal
  field for all galaxies in the LM sample at $z=1$ as a function of environmental density determined
  on scales of  physical $2\,h^{-1}{\rm Mpc}$. Light gray points indicate the data, dark gray points
  show binned median values with errorbars representing the uncertainty on the median. The lines
  represent linear least square fits to the data.}
\end{figure*}

In Figure \ref{fig:delta00257}, we show the dependence on environmental density of the alignment
angles of the galaxies' stellar angular momentum with the three principal axes of the tidal field
for the low-mass sample of galaxies at redshift $z=1$. The overdensity field $\delta$ is evaluated
on the scale of $R_s=2\,h^{-1}{\rm Mpc}$ physical. 

We find that the alignment of stellar angular momentum with the intermediate axis decreases
 strongly when the environmental density increases. This is perfectly consistent with our
 the results from Figure \ref{fig:SpinTide00145} that no alignment with the intermediate axis 
 is present for haloes in regions above the median environmental overdensity. On the other hand, 
 the alignments with the first and the third axis increase with the overdensity of the environment.

All three relations between $\delta$ and the three $|\cos \theta_i|$
are significant. In particular, we find Spearman rank correlation coefficients of $0.32$ with 
$2.47\sigma$ significance for $\theta_3$,  $-0.38$ with $2.96\sigma$ significance for $\theta_2$
and $0.24$ with $1.81\sigma$ significance for $\theta_1$.  We find quantitatively identical but
weaker and slightly less significant correlations also at $z=0.5$.

These results strongly suggest that low-mass galaxies are initially aligned with the intermediate principal
axis, as predicted by tidal torque theory, and that this alignment is subsequently destroyed, possibly by 
environmental effects. Since the alignment with the third principal axis increases during this process, it is plausible
that anisotropic galaxy mergers in the filaments or the galaxy's motion along the filament and 
associated ram-pressure effects or a tidal influence on the gas flow accreting onto the galaxy 
\cite[see also][]{Hahn2009} reorient the angular momentum. 
 A first analysis of correlations of the galaxies' mass accretion rates (as a proxy for
mergers) and ram-pressure with alignment strength did not lead to conclusive results. 

Since tidal torque theory predicts that
angular momentum is generated most efficiently at turnaround, it is also plausible that the galaxy's
motion simply leads to a decorrelation of the tidal field it resides in at later times from that it experienced
at turnaround. Galaxies in higher density regions are likely to have moved further. Specifically, the low-mass 
galaxies at $z=1$ have moved from their Lagrangian position (i.e. their centre
of mass position in the initial conditions of the simulation) on average $3.3\pm1.2\,h^{-1}{\rm Mpc}$, comoving,
when they reside in regions with $\delta<\delta_{50}$, while the overdense sample shows 
a mean displacement of $3.9\pm1.2\,h^{-1}{\rm Mpc}$.  We find correlation coefficients between displacement 
and $\theta_2$ of  $-0.25$. Given that there is no ambiguity in smoothing scale when using the displacements, 
it is to be expected that this correlation coefficient should be much larger if decorrelation were the only reason
for the deviation from tidal-torque prediction for the high-density sample.

Given the inconclusive evidence at this stage, we postpone
a thorough investigation of the origin of the observed density dependence to future work.


\section{Comparison with Observations}
\label{sec:galform_observations}
Most observational results for the alignments of central galaxies exist for elliptical galaxies. Early studies
found an alignment of brightest cluster galaxies with the shape of the cluster itself \citep[e.g.][]{Binggeli1982}.
More recently, the alignment of luminous red galaxies (LRGs) has been detected at high significance
in the Sloan Digital Sky Survey(SDSS) \citep{Mandelbaum2006,Hirata2007}. These LRGs
show radial alignment with their surrounding density peaks. These observational detections
of elliptical galaxy alignment are consistent with theoretical studies that demonstrate that
they form through anisotropic mergers along the filaments at whose nodes they will eventually
reside \citep{West1989,Dubinski1998}. Our results indicate that the spin of the most massive disk galaxies
in our samples show a preference for alignment with the third principal axis of the tidal field which
is parallel to the overall direction of the filaments (cf. Figures \ref{fig:SpinTide00145}, \ref{fig:SpinTide00257},
\ref{fig:SpinTide00439}). This thus corresponds to a radial anti-alignment of the disk shape with the filament.
 Since we however excluded the scales of galaxy groups
from our simulations, we can only speculate on their alignment properties. It is nonetheless perfectly
consistent with observational results that the anti-aligned disk galaxies will merge along the 
massive filament with the merger remnant then being aligned with the filament since the orbital
angular momentum of a major merger will outweigh the contribution from the spin angular
momentum of the contributing disk galaxies \citep[see also][]{Dubinski1998}. Our result could
also explain why the alignment signal of massive galaxies seen by \cite{Mandelbaum2006} and
\cite{Hirata2007} decays so rapidly when less luminous galaxies are considered and thus the
fraction of massive disk galaxies increases.

For lower mass disk galaxies,
the observational findings are however less conclusive. Both \cite{Mandelbaum2006} and \cite{Hirata2007} detect no
alignment for $L<L_\ast$ galaxies in SDSS. On the other hand, \cite{Lee2007} claim a detection
of an alignment of 2MASS galaxies with the intermediate principal axis of the reconstructed
tidal field. Such an alignment with the intermediate axis is consistent with an alignment of galaxy
rotation axes with the shells around cosmic voids. The detection of such an alignment of disk
galaxies around voids has been claimed by \cite{Trujillo2006} in the 2dFGRS and SDSS
galaxy surveys. 
Their analysis was recently repeated by \cite{Slosar2009} on the SDSS DR6 who however
cannot reproduce the \cite{Trujillo2006} result. The \cite{Slosar2009} analysis extends
out to a redshift of $\lesssim 0.18$. The strong density dependence of disk galaxy alignment
that we found in our analysis is likely to impact these analyses of volume limited galaxy
samples. We thus believe that our results are consistent with their non-detection.
It is also plausible that the alignment is weaker at low redshifts.


\section{Summary and Conclusions}
\label{sec:galform_summary}
We have used a cosmological hydrodynamic simulation of galaxy formation in a large-scale 
filament to investigate (1) the orientations of 
the angular momentum of the stellar and gaseous disk of central galaxies, as well as the inner part of their dark 
matter halo with respect to the total angular momentum of the host halo; and (2) the orientations of
their angular momentum with the surrounding large-scale structure by computing scale-dependent
alignments with the tidal field eigenvectors. 

The simulation provides us with a sample of  $\sim100$ disk galaxies at $380\,h^{-1}{\rm pc}$ resolution
down to $z=0$,
spanning halo masses between $10^{11}$ and $10^{13}\,h^{-1}{\rm M}_\odot$ and stellar masses
between $7.5\times10^{9}$ and $8\times10^{11}\,h^{-1}{\rm M}_\odot$. We focus this first analysis on the
three simulation snapshots at redshifts $0$, $0.5$ and $1$. We split our sample of galaxies into 
low-mass galaxies, which are those with a halo mass below the cold accretion limit of $\sim4\times 10^{11}\,h^{-1}{\rm M}_\odot$,
and high-mass galaxies above this limit. At $z=0$, we consider an additional medium mass bin
bridging the gap between the cold accretion limit and the non-linear mass scale of $\sim 2\times10^{12}\,h^{-1}{\rm M}_\odot$.
Our main results are summarised as follows:

\begin{itemize}
\item There is an almost perfect alignment at a median of $\sim 18$ degrees of the stellar, gaseous and inner dark matter angular momenta at
low redshifts. At $z=1$, there is a slightly weaker alignment at $\sim 36$ degrees of the stellar and gaseous 
spins with the dark matter spin, likely due to the higher fraction of unrelaxed galaxies at that epoch. We do not
find any dependence of this alignment signal either on environmental density, or halo mass or stellar mass.

\item The distribution of angles between the spin of the central galaxy and the entire host halo is 
significantly broader, the corresponding median angles larger. We find a median angle of $\sim 50$
degrees between both the stellar and the gas disk and the total halo angular momentum at $z=0$.
This median angle decreases slightly at higher redshifts to a median of $\sim46$ degrees at $z=1$.
The spin of the inner dark matter halo is slightly stronger aligned with the total halo. We find a median
of $\sim 45$ degrees at $z=0$ and $\sim 34$ degrees at $z=1$. Again, there is no dependence
of this alignment signal  on either stellar or halo mass, or on environmental density.

\item Low-mass galaxies in low density regions are aligned with the intermediate
axis of the large-scale tidal field tensor, peaking on scales of physical $2\,h^{-1}{\rm Mpc}$ 
(i.e. $4\,h^{-1}{\rm Mpc}$ comoving) at $z=1$ with very high significance. Such an alignment 
is consistent with the predictions from linear tidal torque theory.

\item In density regions above the median overdensity, on scales of physical $2\,h^{-1}{\rm Mpc}$, we find however 
no evidence for alignment of the  low-mass galaxies with the intermediate principal axis of
the tidal field -- at any redshift since $z=1$. Instead, we find alignment with
the first principal axis at $z=1$, which disappears at later epochs.

\item We find a strong and significant correlation between environmental density and the degree 
of alignment with the intermediate axis at $z=1$. Low mass galaxies gradually misalign with the
intermediate principal axis of the tidal field with increasing environmental density. A possible
origin of this effect could be either torques exerted by ram-pressure when the galaxies enter a 
high-pressure environment,  a change in accretion/merger rates in these regions, a gradual
decorrelation from the tidal field at formation due to the galaxy's motion in dense regions, or a combination 
of these effects. We will quantitatively investigate this issue in a future study.

\item The alignment with the intermediate principal axis of low-mass galaxies in low-density regions 
weakens with decreasing redshift. The interpretation of this result is  not straightforward, as low mass galaxies 
show  some spurious alignment with the AMR mesh at low redshifts. It is important to stress however that we have
evidence, from the analysis of  the $z=1$ and $z=0.5$ snapshots, that non-linear effects do weaken the correlation, 
raising the question as to whether the absence of alignment at $z=0$ is a spurious effect due to numerical effects 
in our simulations, or rather an important physical effect. Simulations are planned which are better tailored  
to establish whether the alignment with the intermediate axis persists in low density regions down to $z=0$. 
Simulations of  isolated galaxies hint at the possibility that an alignment signal could persist down to $z=0$ 
\citep{Navarro04}.

\item Within the caveats of a limited statistical significance, due to small number statistics, an important trend is 
unveiled, for the most massive disk galaxies at all redshifts to be aligned with the third principal axis of the tidal 
field (pointing along the filaments).
\end{itemize}

Summarising, low-mass galaxies  with halo mass below $4\times10^{11}\,h^{-1}{\rm Mpc}$
are likely to show strong alignment with the intermediate principal axis of the tidal field at redshifts relevant 
for weak lensing studies. This alignment is significantly weakened in regions of high environmental density, 
likely due to non-linear effects. We are studying the implications of such alignments for weak lensing experiments 
such as the ESA mission Euclid, under study for launch before the end of the decade to start the quest to 
constrain the nature of Dark Energy.


\section*{Acknowledgments}
OH is grateful for useful discussions with Tom Abel. OH was supported by the Swiss National Science Foundation
during the preparation of this work. All simulations were performed on the Cray XT-3 cluster at CSCS, Manno, Switzerland.




\appendix

\section{Assessment of Numerical Effects on Galaxy Orientations}
\label{sec:numerical_artefacts}
\subsection{The Lower Resolution Simulation}
We also performed a simulation of the filament described in Section \ref{sec:filament_sim} at identical mass resolution 
but with a lower maximum refinement level so that the maximum spatial resolution is $0.76\,h^{-1}{\rm kpc}$ at all
times, a factor of 2 lower than the simulation used in our analysis. We analysed this simulation in an
identical way as the higher resolution one and quantify the differences in what follows in order to
determine the numerical reliability of our results. The lower spatial resolution should especially
demonstrate whether the galaxy orientations are affected by force anisotropy at the resolution level.

We decide to keep the mass samples identical to those in the previous sections so that galaxies on
average contain 8 times less star particles (and thus more than 3000) but an equal 
amount of dark matter particles. We compare our results at $z=0$ as spurious grid alignment effects
are expected to be strongest in the dark energy dominated period where the merger rates decline 
substantially and the galaxies thus could potentially relax to the grid. The low-resolution simulation
contains slightly more galaxies since additional refinement, cooling and star formation were
not restricted to not contain the massive groups. These group galaxies are however not included in our
comparison as we restrict ourselves to the central galaxies of haloes that are resolved in 
both simulations.

\subsection{Angular Momentum Correlations in the Galaxy and with the Host Halo}
\begin{table}
\begin{center}
\begin{tabular}{ll}
\hline
{median} & $z=0$  \\
\hline
$\cos\theta_{\rm S,G}$    & $0.988^{+0.001}_{-0.002}$ \\
$\cos\theta_{\rm S,DM}$ & $0.953^{+0.002}_{-0.019}$ \\
$\cos\theta_{\rm G,DM}$ & $0.924^{+0.004}_{-0.022}$ \\
\hline
$\cos\theta_{\rm S,tot}$    & $0.627^{+0.018}_{-0.055}$ \\
$\cos\theta_{\rm G,tot}$    & $0.594^{+0.013}_{-0.052}$  \\
$\cos\theta_{\rm DM,tot}$ & $0.745^{+0.013}_{-0.048}$ \\
\hline
\end{tabular}
\end{center}
\vspace{0.5cm}
\caption{\label{tab:GalaxyComponentAlignment_lowres}Same as Table \ref{tab:GalaxyComponentAlignment} but for the 
simulation with two-times lower spatial resolution. }
\end{table}

In Table \ref{tab:GalaxyComponentAlignment_lowres}, we give the median angles between the stellar, gas and 
dark matter component within the galaxies and with the total halo angular momentum at $z=0$ for the low-resolution
run. Comparing these angles with those of the high-resolution run in Table \ref{tab:GalaxyComponentAlignment},
we find that the median alignment angles obtained at lower spatial resolution are marginally larger, the
corresponding vectors thus less aligned. Given the width of
the distributions, the results can however be considered consistent at high significance.

\subsection{Isotropy of the Galaxy Spins}
\begin{figure}
	\begin{center}
	  \includegraphics[width=0.3\textwidth]{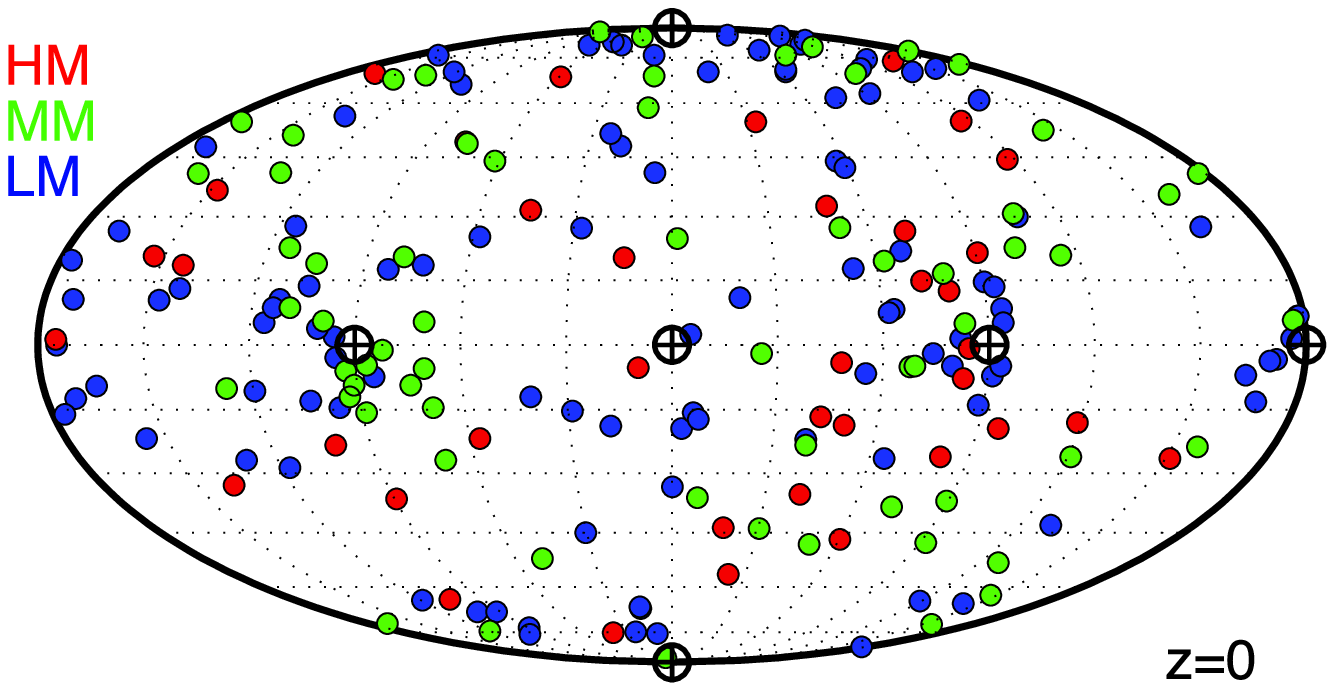}
	\end{center}
	\caption{\label{fig:ProjSpins_lowres}Same as Figure \ref{fig:ProjSpins} but for the low-resolution run at $z=0$.}
\end{figure}

\begin{figure*}
  \begin{center}
  \includegraphics[width=0.9\textwidth]{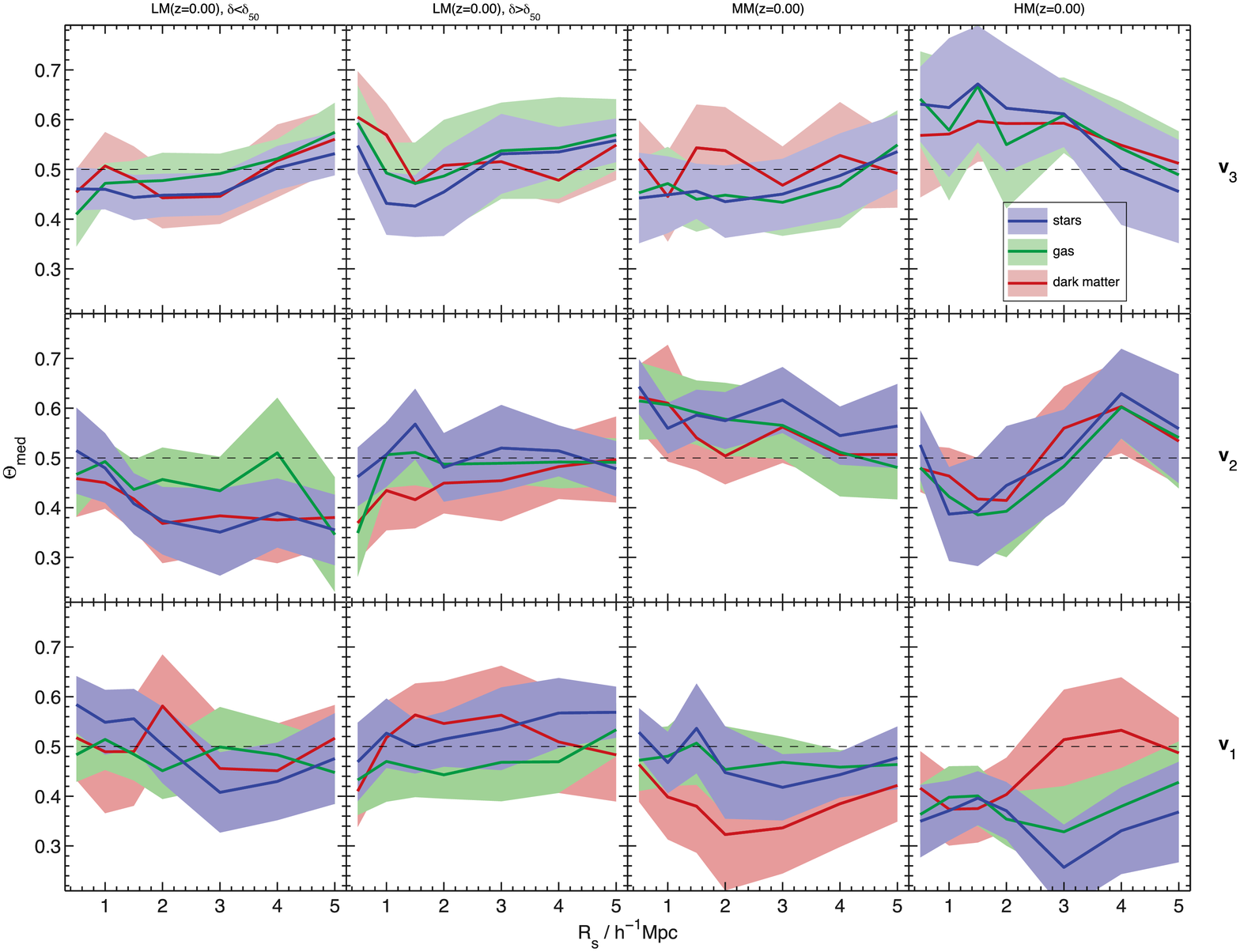}
  \end{center}
  \caption{\label{fig:SpinTide_lowres}Same as Figure \ref{fig:SpinTide00145} but for the simulation with
  two-times lower spatial resolution at redshift $z=0$. The corresponding results for the high-resolution
  run are shown in Figure \ref{fig:SpinTide00439}.}
\end{figure*}

In Figure \ref{fig:ProjSpins_lowres}, we show the orientations of stellar angular momentum vectors for
the low-resolution simulation. This should be compared with the corresponding Figure \ref{fig:ProjSpins} 
for the high-resolution run. Visually, we see a stronger clustering of spin vectors around the
Cartesian mesh basis vectors than for the high-resolution run. Spurious grid alignment is thus present at $z=0$
in both simulations to some degree.

\subsection{Alignments with the Tidal Field}
In Figure \ref{fig:SpinTide_lowres} we show the scale dependence of the alignment between the galaxy stellar,
gas and dark matter angular momentum and the three eigenvectors of the tidal field for the low-resolution
run. These results should be compared with results for the high-resolution simulation given in Figure
\ref{fig:SpinTide00439}.
We find that the results are perfectly consistent for all samples except the low-density low-mass bin
where we find a slightly stronger anti-alignment with the intermediate principal axis of the tidal field
in the low-resolution run. Also, the weak anti-alignment with the third principal axis is not seen in 
the high-resolution run. It is likely that this difference for low-mass galaxies in low
density regions is due to the anisotropy induced by the Cartesian mesh which is expected to
be stronger for the low-resolution run. We conclude that the resolution of our high resolution run is
still insufficient for galaxy orientations to be not affected by spurious grid relaxation in low density regions
at low redshifts. It is thus not clear whether the alignment
of spins with the intermediate axis that we find at higher redshifts would persist in low-density environments also 
down to $z=0$ in the absence of numerical effects. This question has to be addressed with even
higher resolution simulations than the ones we used in our analysis. The results of \cite{Navarro04},
who find a survival of alignment with the intermediate axis for the four isolated galaxies they
simulated, is however indicative that such alignment should persist to some degree.

\label{lastpage}
\end{document}